\documentstyle[prd,aps,preprint,titlepage,epsf,floats]{revtex}
\draft
\tighten

\renewcommand{\epsfsize}[2]{%
   \ifnum\epsfxsize=0
      \ifnum\epsfysize=0
         0.45#1%
      \else
         \epsfxsize
      \fi
   \else
      \epsfxsize
   \fi
}

\renewcommand{\slash}{\not\!}

\begin{document}

\preprint{
        \parbox{1.5in}{%
           \noindent
           hep-ph/9611433 \\
           CERN-TH/96-314\\
           PSU/TH/168
        }
}

\title{Factorization for hard exclusive electroproduction of
       mesons in QCD}

\author{John C. Collins$^{a}$,
        Leonid Frankfurt$^{b}$\thanks{On leave of absence from:
           St.\ Petersburg Nuclear Physics
           Institute, Gatchina, Russia.  },
        Mark Strikman$^{a}$
        }

\address{
        $^{a}$Department of Physics, Penn State University,
            104 Davey Lab., University Park, PA 16802, U.S.A.
\\
        $^{b}$Physics Department, Tel-Aviv University, Tel Aviv,
           Israel
}

\date{July 28, 1997}

\maketitle

\begin{abstract}
    We formulate and prove a QCD factorization theorem for hard
    exclusive electroproduction of mesons in QCD.  The proof is
    valid for the leading power in $Q$ and all logarithms. This
    generalizes previous work on vector meson production in the
    diffractive region of small $x$. The amplitude is expressed
    in terms of off-diagonal generalizations of the usual parton
    densities. The full theorem applies to all kinds of meson and
    not just to vector mesons. The parton densities used include
    not only the ordinary parton density, but also the helicity
    density ($g_{1}$ or $\Delta q$) and the transversity density ($h_{1}$ or
    $\delta q$), and these can be probed by measuring the polarization
    of the produced mesons with unpolarized protons.
\end{abstract}




\section{Introduction}
\label{sec:introduction}

In two recent papers \cite{BFGMS,FKS} it was shown how
the cross section for diffractive electroproduction of vector
mesons can be predicted in perturbative QCD.\footnote{
   Ryskin \cite{Ryskin} considered the case of $J/\psi $ production,
   i.e., that the vector meson is composed of a heavy
   quark and antiquark.  This work used a charmonium model for
   the meson, rather than treating the meson more generally in
   terms of the light-cone wave function that the factorization
   theorem requires.
}
This process provides a novel probe of the dynamics of
diffractive scattering in QCD. One notable prediction is that the
cross section is proportional to the square of the gluon density
in the hadron. Experimental data \cite{ZEUS,H1,ZEUS1} appear to
be in accord with the predictions, including an enhancement due
to the rapid rise of the gluon density at small $x$.

In this paper we extend the factorization theorem to the general
case of electroproduction of any meson, and we provide a
general proof of the theorem.  The theorem expresses the
amplitude for the process in terms of off-diagonal
generalizations of the usual parton densities.
Our demonstration is valid for the whole leading power
for the process, in contrast with the calculations in
Ref.\ \onlinecite{BFGMS,FKS}, which were in a leading-logarithm
approximation.\footnote{
   Ryskin {\em et al.}\ \cite{Ryskin.et.al} treat some of the
   nonleading-logarithm approximation (NLLA) corrections in
   the case of $J/\psi $ production.  In this paper we treat
   very generally corrections to all orders.}
Our results will enable the process to be treated with the
inclusion of non-leading logarithms.  Apart from establishing the
factorization theorem for the process, one aim of this paper is
to attempt a pedagogical exposition of the methods by which the
theorem is derived, since many of the concepts are unfamiliar.

Most importantly, the process of constructing a proof led us to
new results.  First, the theorem applies to the general case of
two-body final states at low transverse momentum in
electroproduction at large $Q$. The diffractive case simply
corresponds to the small-$x$ region, with vacuum quantum number
exchange. So we have extended the theorem to the full range of
$x$ and to all mesons, pions in particular, not just vector
mesons. In addition we find that we need not only the usual
unpolarized parton densities (generalized to be off-diagonal),
but also the helicity densities ($g_{1}$ or $\Delta q$) and the
transversity densities ($h_{1}$ or $\delta q$).  Since the cross section
is proportional to the square of the densities, it is sensitive
to the polarized parton densities without needing a polarized
proton beam and without needing a measurement of the polarization
of the final-state proton.  Indeed we can choose which kind of
density is probed merely by choosing the final state meson. The
amplitude for longitudinally polarized vector mesons depends only
on the unpolarized parton densities. The amplitude for
transversely polarized vector mesons depends only on the
transversity densities ($h_{1}$). The amplitude for pseudoscalar
mesons depends only on the helicity densities.

This result clearly adds to the meager list of processes where
the transversity of valence quarks can be probed without the need
of some other unknown quantity (such as an antiquark density or a
polarized fragmentation function).

All the above statements apply when the incoming virtual photon
is longitudinally polarized.  We also prove that the cross
section is suppressed by a power of $Q$ when the photon is
transversely polarized.

We give a fairly detailed account of the proof of the
factorization theorem.  The style of proof is based on that of
Refs.\ \onlinecite{fact1,fact2,Sterman1,fact3},
which treat inclusive hard scattering.
There are some differences.
First, our derivation of the power-counting formula
shows some useful improvements. Secondly, and rather importantly,
we have to examine more closely exchanges of relatively soft
quarks, since, in contrast to the case of inclusive scattering,
there can be leading contributions from soft quark exchange
(otherwise known as the endpoint contributions). Thus we have to
examine the power-counting arguments in more detail.

After this work was substantially complete, Radyushkin
\cite{Radyushkin2} published a preprint treating some of the
same processes that we consider. His work appears to be
completely compatible with ours; he takes the same point of view
as we do concerning a generalized operator product expansion,
although the details of his notation are a little different.
However, he considered only the diffractive limit of small $x$
for vector meson production, and
hence, just as in Ref.\ \onlinecite{BFGMS}, he did not include
the quark contribution.
(The quark operator
is presumably unimportant at small $x$.)
He does not present a complete proof of factorization.
Ji \cite{Ji} and
Radyushkin \cite{Radyushkin1} also showed how the same operators
appear in an expansion for deeply virtual Compton scattering.

In a future paper we hope to explore further consequences of our
results, including detailed calculations.


\section{Definition of process}
\label{sec:kinematics}

The process we treat is the diffractive exclusive production of
mesons in deep-inelastic electroproduction.  We can
express the lepto-production cross section in terms of the cross
section for the scattering of virtual photons:
\begin{equation}
     \gamma ^{*}(q) + p \to  V(q+\Delta ) + p'(p-\Delta ) .
\label{process}
\end{equation}
The target, of momentum $p^{\mu }$, can
be a proton or nucleus (or any other hadron), and the diffracted
hadron $p'$,
of momentum $p-\Delta $,
may or may not have the same flavor quantum
numbers as the incoming hadron $p$.
The other final-state particle, $V$, can be any
possible meson, e.g., $\rho ^{0}$, $\omega $, $J/\psi $, $\Upsilon $ or $\pi $.
When we treat charge exchange scattering within
our framework,
the direct connection to the parton densities in the proton
\cite{BFGMS,FKS,Ryskin,Ryskin.et.al} is lost.
We will assume that the meson has quantum numbers such that it
cannot decay to a gluon pair.  This choice will eliminate
certain subprocesses, and covers the mesons of interest.

The process depends on three kinematic variables: the virtual
photon's virtuality, $Q^{2}\equiv -q^{2}$, the square of the center-of-mass
energy, $s$ (for the photon-proton system), and the momentum
transfer squared, $t=\Delta ^{2}\leq 0$.  The region we consider is where
$Q^{2} \gg \Lambda ^{2}_{\rm QCD}$, while $|\Delta ^{2}|$ is small,
of order $\Lambda ^{2}_{\rm QCD}$.
We also assume that the meson mass obeys $m_{V}^{2} \ll s$.  We
are thus treating the asymptotics as $Q$ gets large.
The Bjorken variable is $x=Q^{2}/2p\cdot q\approx Q^{2}/(s+Q^{2})$
(where the target mass is neglected).
In Refs.\ \onlinecite{BFGMS,FKS} the diffractive case $x \ll 1$ was
treated. Our considerations will apply to large $x$ as well.

We will mostly restrict our attention to the case that the
virtual photon is longitudinally polarized.  The cross section
with transversely polarized photons is somewhat smaller --- this
was a prediction of Refs.\ \onlinecite{BFGMS,FKS}, and is confirmed
experimentally,\footnote{
   Dominance of production of longitudinally polarized
   mesons has been predicted also by
   Donnachie and Landshoff \cite{Landshoff}
   within a nonperturbative model of the Pomeron.
   This is presumably because their diagrams have to obey the same
   power counting rules as we derive.
}
although the suppression is not as much as one might expect.
Indeed, we find we can derive a simple factorization theorem
only for longitudinally polarized photons, since then the
contributions from the endpoints $z\to 0$ and $z\to 1$ of the meson
wave function are power suppressed,
given that
the meson wave function at its endpoints behaves approximately as
$z(1-z)$.
For transverse polarization,
this suppression does not happen, and a more complicated theorem is
needed---see Sec.\ \ref{sec:transverse}.
At high enough $Q$, there is a Sudakov suppression, but the
physics of this goes beyond the simple factorization theorem,
just as in the analogous case of the electromagnetic form factor
of the proton \cite{Duncan.Mueller}.

It is convenient to use light-front coordinates defined with
respect to the collision axis: $v^{\mu }=(v^{+},v^{-},v_{\perp })$, with
$v^{\pm }=(v^{0}\pm v^{3})/\sqrt 2$.  Then we can write
\begin{eqnarray}
   p^{\mu } &=& \left( p^{+}, \frac {m^{2}}{2p^{+}}, {\bf 0}_{\perp } \right),
\nonumber\\
   q^{\mu } &\approx & \left( -xp^{+}, \frac {Q^{2}}{2xp^{+}},
                {\bf 0}_{\perp } \right),
\nonumber\\
   \Delta ^{\mu } &\approx & \left( xp^{+},
                -\frac {\Delta _{\perp }^{2}+m^{2}x}{2(1-x)p^{+}},
                {\bf \Delta }_{\perp } \right),
\nonumber\\
   V^{\mu } &\approx & \left(
                \frac {\Delta _{\perp }^{2}+m_{V}^{2}}{Q^{2}} xp^{+},
                \frac {Q^{2}}{2xp^{+}},
                {\bf \Delta }_{\perp } \right).
\label{momenta}
\end{eqnarray}
Here, $V^{\mu }$ is the momentum of the meson.  In these
equations, we have neglected small terms in the
longitudinal components, of relative size $\Lambda _{\rm QCD}^{2}/Q^{2}$.
These
coordinates agree with the ones used in Refs.\
\onlinecite{fact1,fact3}, but differ from those in Refs.\
\onlinecite{BFGMS,FKS} by a factor of $\sqrt 2$, and by a
change of the use
of the $+$ and $-$ labels:
$v^{+}_{\rm this\ paper} = {v_{-}}_{\rm Ref.\ \onlinecite{BFGMS}}/\sqrt 2$, and
similarly for $v^{-}$.


\section{Statement of theorem}
\label{sec:theorem}


\subsection{Theorem}

\begin{figure}
    \begin{center}
        \leavevmode
        \epsfxsize=0.4\hsize
        \epsfbox{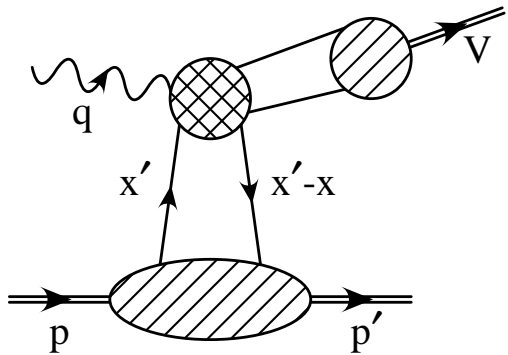}
    \end{center}
\caption{Factorization theorem.}
\label{fig:fact}
\end{figure}

The theorem we will prove is that the amplitude for the process
Eq.~(\ref{process}) is \cite{BFGMS}
\begin{eqnarray}
   {\cal M} &=&
   \sum _{i,j} \int _{0}^{1}dz  \int dx_{1}
   f_{i/p}(x_{1},x_{1}-x,t,\mu ) \,
   H_{ij}(Q^{2}x_{1}/x,Q^{2},z,\mu )
   \, \phi ^{V}_{j}(z,\mu )
\nonumber\\
&&
   + \mbox{power-suppressed corrections} .
\label{factorization}
\end{eqnarray}
Here, $f_{i/p}$ is just like the distribution function for partons
of type $i$ in hadron $p$, except that it is a non-forward matrix
element.\footnote{\label{gen.thm}
    In fact, our whole paper applies to a more general case.  The
    final-state proton in Eq.\ (\ref{process}) may be replaced by
    a general baryon: a neutron, for example.  Then the exchanged
    object no longer has to have vacuum quantum numbers.  The
    index $i$ in the factorization theorem is then to be replaced
    by a pair of indices for the flavors of the two quark lines
    joining the parton density $f_{i/p}$ to the hard scattering.
    Similarly, the two quark lines entering the meson may be
    different, and the index $j$ is to be replaced by a pair of
    indices.
}
We will give the definition later.  The factor
$\phi ^{V}_{j}$ is the light-cone wave function for the meson, and
$H_{ij}$ is the hard scattering function.  The sums are over the
parton types $i$ and $j$ that connect the hard scattering to the
distribution function and to the meson. Since the meson has
non-zero flavor, the parton $j$ is restricted to be a quark. The
factorization theorem Eq.~(\ref{factorization}) is illustrated in
Fig.\ \ref{fig:fact}.

The above formula is correct for the production of longitudinally
polarized vector mesons.  For the production of transversely
polarized vector mesons or of pseudo-scalar mesons, we have a
formula of exactly the same structure, but in which the
unpolarized parton density is replaced by a polarized parton
density (the transversity density for transverse vector mesons,
and the helicity density for pseudo-scalar mesons).  Similar
changes will need to be made to the definition of the meson wave
function.

The parameter $\mu $ in Eq.~(\ref{factorization}) is the usual
renormalization/factorization scale.  It should be of order
$Q$, in order that the hard scattering function $H_{ij}$ be
calculable by the use of finite-order perturbation
theory.  The $\mu $-dependence of the distribution $f_{i/p}$ and of the
light-cone wave function $\phi _{j}^{V}$ are given by equations of the
Dokshitzer-Gribov-Lipatov-Altarelli-Parisi (DGLAP)
kind, as we will discuss in Sec.\ \ref{sec:evolution}.

\begin{figure}
    \begin{center}
        \begin{tabular}{c@{~~}c}
           \epsfxsize=0.20\hsize
           \epsfbox{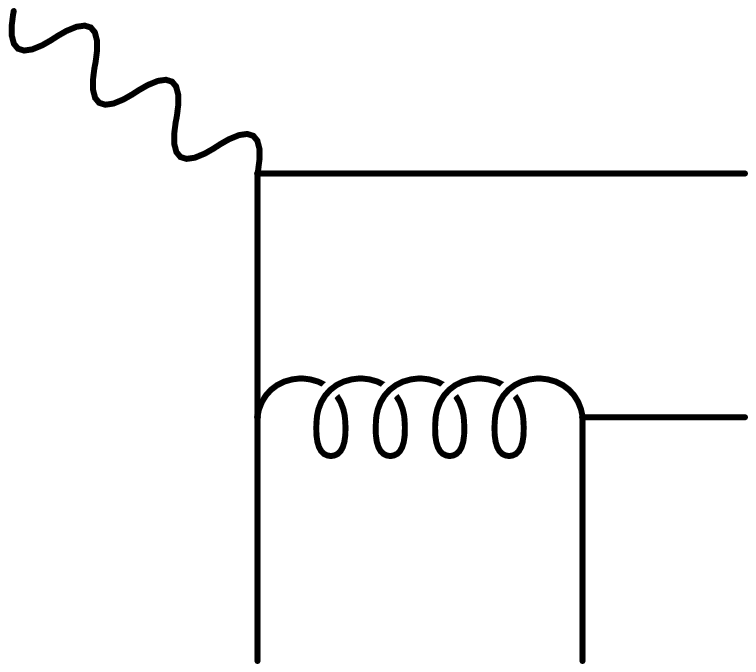}
           \hspace*{0.2in}
           &
           \epsfxsize=0.20\hsize
           \epsfbox{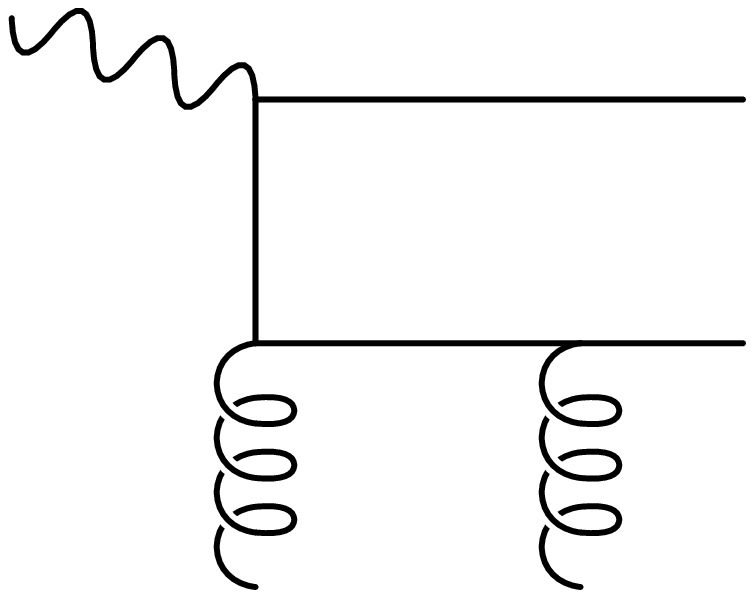}
           \hspace*{0.2in}
        \\
           (a) & (b)
        \end{tabular}
    \end{center}
\caption{Typical lowest-order graphs for $H$.}
\label{fig:LO.Graphs}
\end{figure}

Typical lowest order graphs for $H$ are shown in Fig.\
\ref{fig:LO.Graphs}.  Consider first graph (a), all of whose
external lines are quarks. After we go through the derivation of
the factorization theorem, and have constructed definitions of
the distribution $f_{i/p}$ and of the light-cone wave function $\phi ^{V}$,
we will be able to see that the definition of $H$ is the sum of
graphs such as Fig.\ \ref{fig:LO.Graphs}(a) contracted with suitable
external line factors that correspond to the Dirac wave functions of
the partons.  In the case of longitudinal vector meson
production, the factors are $\frac {1}{2}p^{+}\gamma ^{-}$
for the lower two lines and $\frac {1}{2}V^{-}\gamma ^{+}$ for the
lines connected to the outgoing meson. These factors are related
to spin averages of Dirac wave functions for the quarks.

In the case of the gluon-induced subprocess, Fig.\
\ref{fig:LO.Graphs}(b), the external fermion
lines of $H$ are to be contracted with the same factors as
before, but the two gluon lines are to be contracted with
$\delta ^{\alpha \beta }/2$, where $\alpha $ and $\beta $ are
transverse indices, and the $1/2$
represents a kind of spin average.

See Sec.\ \ref{sec:hard} for more information on the precise
normalization conventions for the hard scattering function.


\subsection{Definitions of light-cone distributions and
            amplitudes: longitudinal vector meson}

\subsubsection{Quark distribution}

The distribution function $f_{i/p}$ and meson amplitude $\phi ^{V}_{j}$
are defined, as usual, as matrix elements of gauge-invariant
bilocal operators on the light-cone.  In the case of a quark of
flavor $i$, we define
\begin{eqnarray}
   f_{i/p}(x_{1},x_{2},t,\mu )  &=&
   \int _{-\infty }^{\infty } \frac {dy^{-}}{4\pi }
   \;
   e^{-ix_{2}p^{+}y^{-}}
   \langle p'|\; T {\bar \psi }(0,y^{-},{\bf 0}_{T})\gamma ^{+}
     {\cal P} \psi (0)\; |p\rangle  ,
\label{pdf.q.def}
\end{eqnarray}
where $\cal P$ is a path-ordered exponential of the gluon field
along the light-like line joining the two operators for a quark
of flavor $i$.
We have defined $x_{1}$ to be the fractional momentum given by the
quark to the hard scattering and $-x_{2}$ to be the momentum given
by the antiquark; in the factorization theorem they obey
$x_{1}-x_{2} = x$, with $x$ being the usual Bjorken variable.
At first sight the right-hand-side of
Eq.~(\ref{pdf.q.def}) appears to depend only on $x_{2}$ and not on
$x_{1}$ nor on $t$.  The dependence on the other two variables comes
from the fact that the matrix element is non-forward.  The
difference in momentum between the states $|p\rangle $ and $|p'\rangle $
together with the use of a light-cone operator brings in
dependence on $x_{1}$ and on $t$.
It is necessary to take only the connected part of the matrix
element.

The same definition has recently been given and discussed by Ji
and Radyushkin \cite{Radyushkin2,Ji,Radyushkin1}.  As Ji points
out, when $t\not=0$ there are in fact two separate parton
densities, with different dependence on the nucleon spin.  For
the purposes of our proof, it will be unnecessary to take this
into account explicitly; we can simply suppose that this and the
other parton densities have dependence on the spin state of the
hadron states $|p\rangle $ and $|p'\rangle $.

The usual quark density $f_{i/p}(x,\mu )$ is obtained by setting
$t=0$ and $x_{1} = x_{2} = x$ in Eq.~(\ref{pdf.q.def}).  In addition,
it would appear that one has to remove the time-ordering
operation from the operator
operators in Eq.\ (\ref{pdf.q.def}) to obtain the
operator used for the parton densities associated with inclusive
scattering \cite{pdfs}.  We need time-ordered operators in our
present work because we are discussing amplitudes
rather than cut amplitudes.
Thus if one sets $t=0$ and $x_{1} = x_{2} = x$ in our parton
distributions, one would naturally suppose that the conventional
inclusive parton densities are the discontinuities of
ours.\footnote{
    Equivalently one would say the the conventional parton densities
    are given by the imaginary part of our distributions.  To be
    precise, with our definitions, which do not possess an
    overall factor of $i$, the discontinuity is twice the real
    part.
}
Relating the new parton densities to the standard ones, even in
the forward limit would therefore appear to need dispersion
relations.

In fact, the two kinds of parton density are equal, at least in
the forward limit.  A proof of this not very obvious fact was
given many years ago by Jaffe \cite{Jaffe}.  However, his proof
applies only to two-particle-irreducible graphs for the parton
densities, a restriction we suspect to be unnecessary.  We hope
to return to this issue in a later paper, particularly because
there are some additional complications in the non-forward parton
densities that particularly appear when one treats dispersion
relations for the amplitude for our process.

It is also worth noting that there is a limit on $t$:
\begin{equation}
    -t > t_{\rm min} = \frac {m^{2}(x_{1}-x_{2})^{2}}{1-x_{1}+x_{2}} ,
\end{equation}
which comes from the kinematics of the scattering proton.
Note that the same limit is obtained from the kinematics of the
scattering process we consider, (\ref{process}), in the limit
$Q \gg m$. We deduce that the limit $t\to 0$ cannot be accessed
directly in exclusive meson production.  Indeed, since
$x_{1}-x_{2}=x_{\rm bj}$, the analytic continuation from $t \not= 0$ to
$t=0$ is hard to perform in practice, except at small $x_{\rm bj}$.

\subsubsection{Gluon distribution}

An exactly similar definition applies for the gluon distribution:
\begin{eqnarray}
   f_{g/p}(x_{1},x_{2},t,\mu )  &=&
   - \int _{-\infty }^{\infty } \frac {dy^{-}}{2\pi }
   \, \frac {1}{x_{1} x_{2} p^{+}}
   \;
   e^{-ix_{2}p^{+}y^{-}}
   \langle p'|\; T G_{\nu }{}^{+}(0,y^{-},{\bf 0}_{T}) \,
     {\cal P} \, G^{\nu +}(0)\; |p\rangle  .
\label{pdf.g.def}
\end{eqnarray}
The $1/x_{1}x_{2}$ factor cancels a inverse factor that appears in the
derivative part of the fields $G_{\nu }{}^{+}(0,y^{-},{\bf 0}_{T}) G^{\nu
+}(0)$.
The normalization is now a little different from that of the
diagonal distribution:
\begin{equation}
   f_{g~\rm diag}(x) = x f_{g~\rm non-diag}(x,x),
\end{equation}
i.e., one sets $x_{1}=x_{2}=x$, and puts in a factor $x$.  To avoid
this complication while preserving symmetry between the two gluon
lines would involve square root factors, or changing the hard
scattering formula Eq.\ (\ref{factorization}) when the partons
are gluons.  The square roots are undesirable, because they
change the analyticity properties of the formula in the
neighborhood of $x_{1}=0$ and $x_{2}=0$.

\subsubsection{Wave function}

The light-cone wave function for a longitudinally
polarized vector meson is \cite{Brodsky.etal.elastic}
\begin{eqnarray}
   \phi ^{V}_{j}(z,\mu ^{2})  &=&
   \frac {1}{\sqrt {2N_{c}}}
   \int _{-\infty }^{\infty } \frac {dy^{+}}{4\pi }
   \;
   e^{-izV^{-}y^{+}} \langle 0|\; {\bar \psi }(y^{+}, 0, {\bf 0}_{T})\gamma
^{-}
  {\cal P} \psi (0) \; |V\rangle  ,
\label{wf.def}
\end{eqnarray}
where the factor of $1/\sqrt {2N_{c}}$ is the convention established by
Brodsky and Lepage \cite{Brodsky.Lepage}---see their Eq.\
(64).  This convention results in a elegant normalization
condition for light-cone wave functions, Eq.\ (26) of Ref.\
\onlinecite{Brodsky.Lepage}.

Our definition appears to disagree with theirs, but
this is fact not so.
We have an extra overall factor $2$ which
merely results from the $1/\sqrt 2$'s in our definition of light-cone
coordinates.  We are missing a $\gamma _{5}$ that they have, because our
meson is a vector instead of a pseudoscalar, and we therefore
need a different operator to pick out the nonzero component.  In
addition, we have exchanged the use of the $+$ and $-$ components
of vectors.  This simply corresponds to the fact that we wish to
apply the definition to a meson that travels in the $-z$
direction in our coordinate system.
The factor of $P_{\pi }^{+}$ in Brodsky and Lepage's definition is an
error, and should be omitted \cite{Brodsky.pc}: their definition
is not invariant under boosts in the $z$ direction.

All of the above definitions have ultra-violet divergences.  So
they are defined \cite{pdfs} to be renormalized by some suitable
prescription, of which minimal subtraction is the standard one.
We do not explicitly indicate the renormalization, which is done
by a factor convoluted with the right-hand sides of these
definitions. The scale associated with the renormalization is
$\mu $, and the DGLAP evolution equations are the
renormalization-group equations for the $\mu $ dependence.

As stated in footnote \ref{gen.thm}, there is a more general
theorem, in which the final-state hadron in the distribution Eq.\
(\ref{pdf.q.def}) has quantum numbers different from the proton.
Then it would be necessary to modify this definition, so that the
quark and antiquark fields have different flavors.  (The gluon
distribution would also be zero.)  Similar modifications would be
needed to the meson amplitude Eq.\ (\ref{wf.def}).


\subsection{Definitions of light-cone distributions and
            amplitudes: pseudo-scalar meson and transverse vector meson}

When we write the factorization formula for a pseudo-scalar
meson, different components components of Dirac matrices dominate
in the amplitudes.  We will see that the following changes are
needed in the definitions, Eqs.\ (\ref{pdf.q.def}),
(\ref{pdf.g.def}) and (\ref{wf.def}):
\begin{equation}
\begin{tabular}{|l|l|l|}
    \hline
    Object              & original      & replacement
                                          (pseudo-scalar meson)
    \\
    \hline
    \hline
    Meson wave function &  $\gamma ^{-}$         & $\gamma ^{-}\gamma _{5}$
    \\
    Quark density       &  $\gamma ^{+}$         & $\gamma ^{+}\gamma _{5}$
    \\
    Gluon density       &  $-G_{\nu }{}^{+} G^{\nu +}$ & Not used
    \\
    Coupling of $H$ to
    quarks from meson   &  $q^{-}\gamma ^{+}/2$     & $q^{-}\gamma _{5}\gamma
^{+}/2$
    \\
    Coupling of $H$ to
    quarks from proton  &  $p^{+}\gamma ^{-}/2$     & $p^{+}\gamma _{5}\gamma
^{-}/2$
    \\
    Coupling of $H$ to
    gluons from proton  &  $\delta ^{ij}/2$     & Not used
    \\
    \hline
\end{tabular}
\label{pseudo.scalar.changes}
\end{equation}
The parton densities in the diagonal limit then correspond to the
helicity densities \cite{CTEQ}
$\Delta f$ that are used in the treatment of the
polarized structure function $g_{1}$.
However, the gluon density is not used: charge conjugation
invariance implies that the hard scattering coefficient is zero
when it couples a virtual photon and a pseudo-scalar meson to a
pair of gluons.

For a transversely polarized vector meson, we use the following
replacements
\begin{equation}
\begin{tabular}{|l|l|l|}
    \hline
    Object              & original      & replacement
                                          (transverse vector meson)
    \\
    \hline
    \hline
    Meson wave function &  $\gamma ^{-}$         & $\gamma ^{-}\gamma
^{i}\gamma _{5}$
    \\
    Quark density       &  $\gamma ^{+}$         & $\gamma ^{+}\gamma
^{j}\gamma _{5}$
    \\
    Gluon density       &  $-G_{\nu }{}^{+} G^{\nu +}$ & Not used
    \\
    Coupling of $H$ to
    quarks from meson   &  $q^{-}\gamma ^{+}/2$     & $q^{-}\gamma _{5}\gamma
^{i}\gamma ^{+}/2$
    \\
    Coupling of $H$ to
    quarks from proton  &  $p^{+}\gamma ^{-}/2$     & $p^{+}\gamma _{5}\gamma
^{j}\gamma ^{-}/2$
    \\
    Coupling of $H$ to
    gluons from proton  &  $\delta ^{ij}/2$     & Zero
    \\
    \hline
\end{tabular}
\label{transverse.changes}
\end{equation}
Note that the gluon density does not appear in this case, for
reasons of helicity conservation in the hard scattering. In the
diagonal limit, the quark density we use with transversely
polarized vector mesons becomes the transversity density \cite{CTEQ}
$\delta f_{q}$,
also called $h_{1}$.

The combinations of Dirac matrices in the wave functions for
longitudinal vector mesons and pseudo-scalar mesons pick out
pairs quark and antiquarks that have opposite helicity and hence
of the chirality; this is correct for making a meson of zero
helicity.  In contrast for a transverse vector meson, the quark
and antiquark have the same helicities and the opposite
chiralities.

\subsection{Real and imaginary parts of amplitude}

In the factorization theorem, Eq.\ (\ref{factorization}),
the amplitude for our process at the hadronic level is expressed
in terms of a hard scattering amplitude together with a
generalized parton density in the proton and a light-cone wave
function of the meson.  Now both the hadronic amplitude and the
hard scattering amplitude satisfy dispersion relations that
relate their real and imaginary parts, and it is not entirely
obvious that the dispersion relations for the two amplitudes are
consistent with the factorization theorem.  Moreover, one might
suppose that complications arise because the cut of the amplitude
needed to obtain the discontinuity of the hadronic amplitude must
cut both the hard scattering amplitude and the parton density in
Fig.\ \ref{fig:fact}.

We now demonstrate that the two dispersion relations are in fact
consistent. The proof will be to demonstrate that the dispersion
relation for the hadronic amplitude follows from the
corresponding dispersion relation for the hard scattering
amplitude.  This is important because one of the approaches to
calculations has been to calculate the imaginary part of the
amplitude first and then to use dispersion relations to compute
the full amplitude. A consequence is that the real and imaginary
parts of the hadronic amplitude are separately expressed in terms
of the real and imaginary parts of the hard scattering amplitude
with the same parton densities.

We will find it convenient to write the amplitude as a function
of $\nu \equiv 2p\cdot q=Q^{2}/x$ rather than $s=\nu -Q^{2}$.  We have
${\cal M}={\cal M}(Q^{2}/x,t,Q^{2})$ and $H=H(Q^{2}x_{1}/x,Q^{2},z)$, where
$x_{1}$
is the same variable as in Eq.\ (\ref{factorization}). The
important fact that lets our derivation work is that $H$ depends
on the ratio $x_{1}/x$ but not on $x_{1}$ and $x$ separately. This is
proved by observing that $H$ is invariant under Lorentz boosts in
the $z$ direction and that a change of $x_{1}$ and $x$ by a common
ratio is equivalent to a boost.

The dispersion relation for the hard scattering amplitude is
\begin{eqnarray}
    H(\nu ,Q^{2},z) &=& \int \frac {d\nu '}{2\pi i} \frac {1}{\nu '-\nu } {\rm
disc}\,H(\nu ',Q^{2},z).
\label{DR.H}
\end{eqnarray}
By choosing the contour to run along the real axis, we have made
the right-hand side of this equation involve only the
discontinuity (or imaginary part) of the amplitude.  Any
subtractions needed in the dispersion relation will not affect
the principles of the derivation.

We now substitute Eq.\ (\ref{DR.H}) in the factorization theorem.
Then writing $\nu '=x_{1}\hat \nu $ gives the dispersion relation
for $\cal M$:
\begin{eqnarray}
   {\cal M}(\nu ,Q^{2},t) &=&
       \int dx_{1} dz \frac {d\nu '}{2\pi i} \frac {1}{\nu '-\nu x_{1}} H(\nu
',Q^{2},z)  \phi (z) f(x_{1},x').\
\nonumber\\
    &=& \int dx_{1} dz \frac {d\hat \nu }{2\pi i} \frac {1}{\hat \nu -\nu }
        H(x_{1}\hat \nu ,Q^{2},z)  \phi (z) f(x_{1},x').
\nonumber\\
    &=& \int  \frac {d\hat \nu }{2\pi i} \frac {1}{\hat \nu -\nu }
        {\rm disc}\,{\cal M}(\hat \nu ,Q^{2},z),
\label{DR.M}
\end{eqnarray}
where in the last line, we have used the factorization theorem
again.  This equation is just the expected dispersion relation
for the hadronic amplitude.

The discontinuity of an amplitude is obtained by making a cut
that puts some intermediate states on shell.  The only possible
cut of $\cal M$ in its factorized form Fig.\ \ref{fig:fact} is
one that cuts both the hard scattering amplitude $H$ and the
parton density $f$.  The statement that the parton densities are
the same whether the operators are unordered or are time-ordered
is equivalent to saying that the cut amplitude equals the uncut
amplitude.  This is consistent with our derivation of the
dispersion relation for $\cal M$.


\section{Regions}
\label{sec:regions}

We wish to calculate the asymptotics in a double limit:
$Q/m\to \infty $ and $x\to 0$, but it is the $Q\to \infty $ limit that we will
concentrate on, since that will result in the perturbatively
calculable factors in our theorem.  It will also give us a more
general theorem, that is applicable at large $x$.
In this and the next section we follow the treatment
of Libby and Sterman \cite{Sterman1,LibSt,Sterman} adapted to
our process.

Graphs for the process have
integrals over all their loop momenta, and we wish to classify the
regions of loop momenta in a suitable way for extracting the
asymptotics as $Q\to \infty $.
To expose the powers of $Q$, we choose to work in the Breit frame where
the virtual photon has zero rapidity,
$xp^{+} = Q^{2}/2xp^{+} = Q/\sqrt 2$. \footnote{
  None of our arguments would change if we made a finite boost.
  Then we would have $xp^{+} \sim Q^{2}/xp^{+} \sim Q$.
}
In such a frame the meson $V$ is moving very fast in one
direction, and the incoming and outgoing protons are moving very
fast in the opposite direction.  The steps in the proof are as
follows:
\begin{enumerate}
\item Scale all momenta by a factor $Q/m$, so that we are in
   effect attempting to take a massless on-shell limit of the
   amplitude.

\item Use the Coleman-Norton theorem to locate all pinch-singular
   surfaces in the space of loop integration momenta, in the
   zero-mass limit.

\item Identify the relevant regions of integration as
   neighborhoods of these pinch singular surfaces.

\item The scattering amplitude is a sum of contributions, one for
   each pinch singular surface, plus a term where all lines have
   virtuality of at least of order $Q^{2}$.  Appropriate
   subtractions are made to prevent double counting.

\item Perform power counting to determine which regions give
   the largest power of $Q$.

\item Finally, show that the contributions for the leading
   power of $Q$ give the factorization formula
   Eq.~(\ref{factorization}).

\end{enumerate}
Any terms that do not contribute to the leading power are
dropped.  The factorization formula is intended to include all
logarithmic corrections to the leading power, whether they are
leading or non-leading logarithms.


\subsection{Scaling of momenta}

Following Libby and Sterman\cite{LibSt} we write a general
momentum $k^{\mu }$ and a general mass $m$ in units of the large
momentum scale $Q$:
\begin{equation}
   k^{\mu }  = Q \tilde k ^{\mu },  ~~~  m  = Q \tilde m.
\end{equation}
Since we work in the rest frame of the virtual photon, i.e., in
the Breit frame, both of the light-cone components of its
momentum are of order $Q$. When everything is expressed in terms
of the scaled variables, $\tilde k$ and $\tilde m$, simple
dimensional analysis shows that the large-$Q$ limit is equivalent
to a zero-mass limit, $\tilde m \to  0$.  Since the amplitude $\cal
M$ is dimensionless, we have
\begin{equation}
   {\cal M}(Q^{2}; p, p_{V}, \Delta , m; \mu ) =
   {\cal M}(1; \tilde p, \tilde p_{V}, \tilde\Delta , \tilde m; \mu /Q) ,
\end{equation}
by ordinary dimensional analysis. Notice that in the limit $Q\to \infty $
\begin{eqnarray}
   \tilde p ^{\mu }  &\to & (p^{+}/Q, 0, {\bf 0}_{\perp }) ,
\nonumber\\
   \tilde q ^{\mu }  &\to & (-xp^{+}/Q, Q/(2xp^{+}), {\bf 0}_{\perp }) ,
\nonumber\\
   \tilde \Delta  ^{\mu }  &\to & 0 ,
\nonumber\\
   \tilde V ^{\mu }  &\to & (0, Q/(2xp^{+}), {\bf 0}_{\perp }) ,
\end{eqnarray}
so that $\tilde p$ and $\tilde V$ become light-like vectors, cf.
Eq.\ (\ref{momenta}).

We consider the most basic region to be where all internal lines
obey $k^{2} \gtrsim Q^{2}$, and thus the scaled momenta $\tilde k$ have
virtualities of order unity, or bigger.  In such a region, we can
legitimately set the mass parameters to zero, and make the
external hadrons light-like.  Most importantly, we will be
entitled to choose the renormalization scale $\mu $ of order $Q$
without obtaining any large logarithms.  Consequently, in this
region an expansion to low order in powers of the small coupling
$\alpha _{s}(Q)$ is useful.

However, this basic region is not the only one.  Indeed, it does
not even provide a leading contribution for the amplitude for our
particular process. But now one observes \cite{LibSt} that all
other relevant regions correspond to singularities of massless
Feynman graphs.  They are neighborhoods of surfaces where the
loop momenta are trapped at singularities, i.e., of
pinch-singular surfaces of the massless graphs.  The conditions
for a pinch singularity are exactly the Landau conditions for a
singularity of a graph.\footnote{
    The relevant singularities are on the physical sheet of the
    space of complex angular momenta, or on its boundary.  Thus
    it is indeed the Landau conditions that are correct.
}
Only pinch singularities are relevant,
since at a non-pinched singularity, we may deform the
(multi-dimensional) integration contour such that at least one
of the singular propagators is no longer near its pole.

If there is a pinch singularity caused by certain propagator
poles in the massless limit, then in the real graph, with nonzero
masses but large $Q$, the contour of integration is forced to
pass near the propagator poles.  Consequently it is not possible
to neglect the masses in this region.  Conversely, if the contour
is not trapped by the poles, then the contour may be deformed
away from the poles, and masses may be neglected in evaluating
the corresponding propagators.


\subsection{Coleman-Norton theorem}

We now review the theorem of Coleman and Norton \cite{CN}, and
show how \cite{Sterman} to apply it. The theorem shows in a
physically appealing fashion how to determine the configurations
of loop momenta that give pinch singularities; it states that
they correspond to classically allowed scattering processes,
treated in coordinate space.

More precisely, the theorem states that each point
on a pinch-singular surface (in loop momentum space) corresponds
to a space-time diagram obtained as follows.  First we obtain a
reduced graph by contracting to points all of the lines whose
denominators are not pinched.  Then we assign space-time points
to each vertex of the reduced graph in such a way that the
pinched lines correspond to classical particles.  That is, to
each line we assign a particle propagating between the space-time
points corresponding to the vertices at its ends.  The momentum
of the particle is exactly the momentum carried by the line,
correctly oriented to have positive energy.
If for some set of momenta, it is not possible to construct such
a reduced graph, then we are free to deform the contour of
integration.

A reduced diagram corresponds to a classically allowed
space-time scattering process.  The construction of the most
general reduced graph becomes extremely simple in the zero mass
limit, since then all pinched lines must carry either a
light-like momentum or zero momentum. Moreover, as was explained
by Libby and Sterman, each light-like momentum must be parallel
to one of the (light-like) external lines.


\subsection{Reduced graphs}

\begin{figure}
    \begin{center}
        \leavevmode
        \epsfxsize=0.25\hsize
        \epsfbox{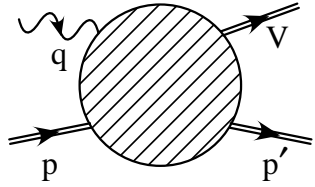}
    \end{center}
\caption{Quasi-elastic scattering of a virtual photon on a
   proton. }
\label{fig:process}
\end{figure}

In the zero mass limit, our process, represented in Fig.\
\ref{fig:process}, has
\begin{itemize}

\item One light-like incoming proton line of momentum
   $p_{A}^{\mu }=(p^{+},0,0_{\perp })$.

\item One light-like outgoing proton line of a slightly
   different, but parallel, momentum
   \\
   $p_{A}'^{\mu }=((1-x)p^{+},0,0_{\perp })$.

\item One light-like outgoing meson line of momentum
   $p_{B}^{\mu }=(0,Q^{2}/2xp^{+},0_{\perp })$.

\item One incoming virtual photon of momentum
   $q^{\mu }=(-xp^{+}, Q^{2}/2xp^{+}, 0_{\perp })$.
\end{itemize}
We have chosen the symbols for the light-like momenta, $p_{A}$,
$p_{A}'$, and $p_{B}$, to be different from the symbols for the
corresponding physical momenta, $p$, $p'$, and $V$, precisely to
emphasize that they are distinct (if related) momenta.

\begin{figure}
    \begin{center}
        \leavevmode
        \epsfxsize=.40\hsize
        \epsfbox{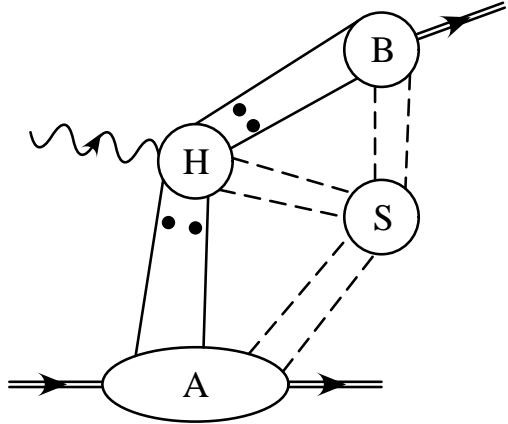}
    \end{center}
\caption{General reduced graph for Fig.\ \protect\ref{fig:process}.
    The dots represent the possibility of an arbitrary number of
    lines connecting the collinear subgraphs ($A$ and $B$) to the
    hard subgraph $H$.  Any number of lines connect the soft
    subgraph $S$ to the other subgraphs.  }
\label{fig:Reduced.Graph}
\end{figure}

As we will prove in Sec.\ \ref{sec:construct.reduced.graphs},
the most general reduced graph is depicted in Fig.\
\ref{fig:Reduced.Graph}.  One vertex of the reduced graph is the
hard subgraph $H$, to which is attached the virtual photon.
The incoming and outgoing protons go into the collinear
subgraph $A$; at the corresponding pinched momentum
configuration, lines in $A$ have only a $+$ component.
Similarly, the outgoing meson is attached to another collinear
subgraph $B$ where there are momenta with only a $-$ component.
Each of the collinear subgraphs is attached to the hard subgraph
by at least one line, and these three subgraphs are all connected;
these restrictions are needed so that momentum conservation works
out.  Finally there may be a soft subgraph, $S$, composed of zero
momentum lines at the pinch singular surface.  It connects to any
of the other subgraphs, and it may have more than one connected
component.

Within any of $A$, $B$ and $S$, there may be subgraphs composed of
hard lines; these form reduced vertices that couple the different
lines within the subgraphs.  In the leading regions, these
are of the form of the possible ultra-violet divergent subgraphs.


\subsection{Space-time interpretation}

\begin{figure}
    \begin{center}
        \leavevmode
        \epsfxsize=1.8in
        \epsfbox{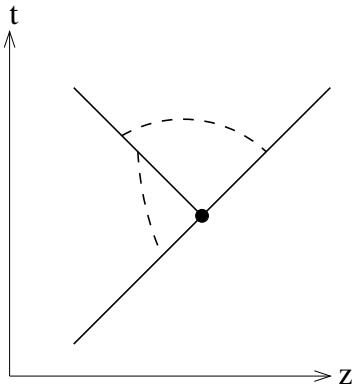}
    \end{center}
\caption{Space-time diagram for Fig.\
         \protect\ref{fig:Reduced.Graph}.}
\label{fig:Space.Time}
\end{figure}

The corresponding space-time diagram is Fig.\
\ref{fig:Space.Time}.  There, each solid line corresponds to a
light-like line of the reduced graph, with a $45^{\circ }$ orientation to
correspond to their light-like lines of propagation.  The dashed
lines correspond to the soft lines, in the subgraph $S$.  From
the point of view of the Coleman-Norton theorem, they are rather
degenerate lines.  Indeed, the fact that they are carrying
zero-momentum (at the singular point) implies that they have no
specific orientation.  Thus we indicate them by curved lines of
no particular orientation. The locations of the endpoints of the
soft lines, where they attach to the collinear subgraphs, can be
anywhere along the world lines of the collinear lines.  The hard
vertex $H$ occurs at the intersection of the collinear lines.

Since there can in general be more than one collinear line moving
in each of the $+$ and $-$ directions, the solid lines in Fig.\
\ref{fig:Space.Time} must each be thought of as a group of lines
which undergo interactions as they propagate.

When the space-time representation of a Feynman graph is used,
there is normally an exponential suppression when there are large
space-time separations between the vertices.  One obtains a
singularity when the exponential suppression fails, and the
Coleman-Norton construction gives exactly the relevant
configurations of the vertices.  A common scaling can be applied
to all the world lines in the reduced graph without affecting its
properties, and the singularity is generated by the possibility
of integrating over arbitrarily large scalings in coordinate
space without obtaining an exponential suppression.

{\em The whole of the discussion above relies on the use of a
covariant gauge.}  Although the use of the axial gauge and in
particular of the light-cone gauge is very convenient, for
example, for a physical interpretation of the light-cone wave
function, the propagators in such a gauge have unphysical
singularities.  The unphysical singularities do not give the
normal rules of causal relativistic propagation of particles,
and, beyond the leading-logarithm approximation, they make the
derivation of the factorization theorem very difficult --- see
\cite{fact1,ColSt}.


\subsection{Examples}

\begin{figure}
    \begin{center}
        \begin{tabular}{c@{~~~~}c}
           \epsfxsize=0.30\hsize
           \epsfbox{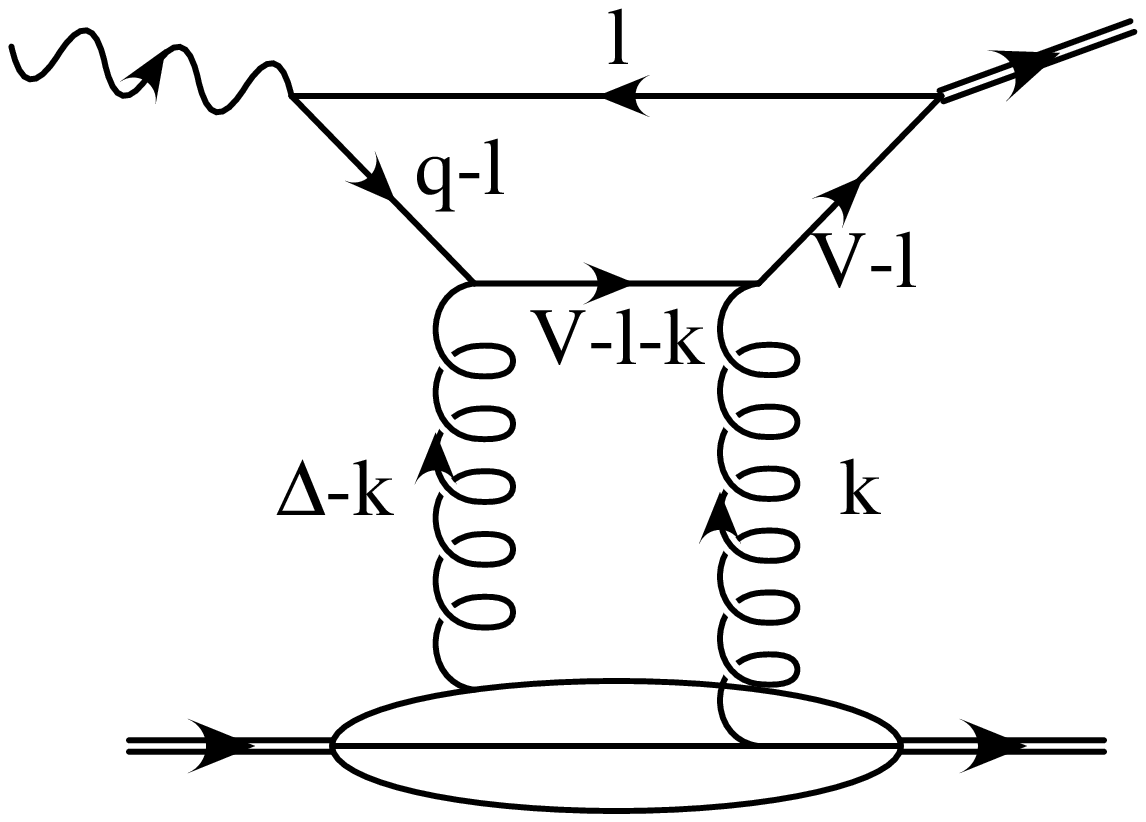}
           &
           \epsfxsize=0.28\hsize
           \epsfbox{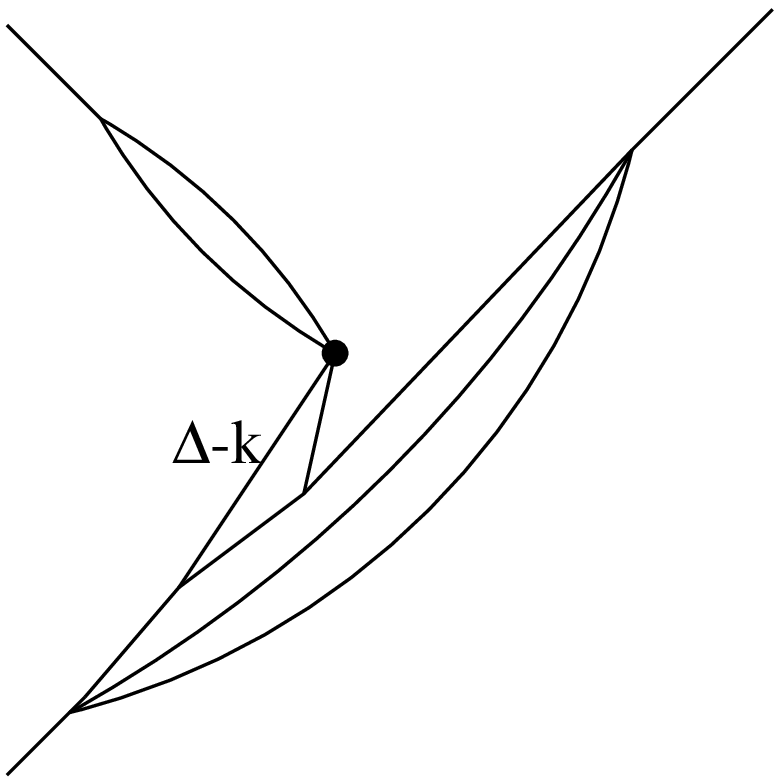}
        \\
           (a) & (b)
        \\
           \epsfxsize=0.28\hsize
           \epsfbox{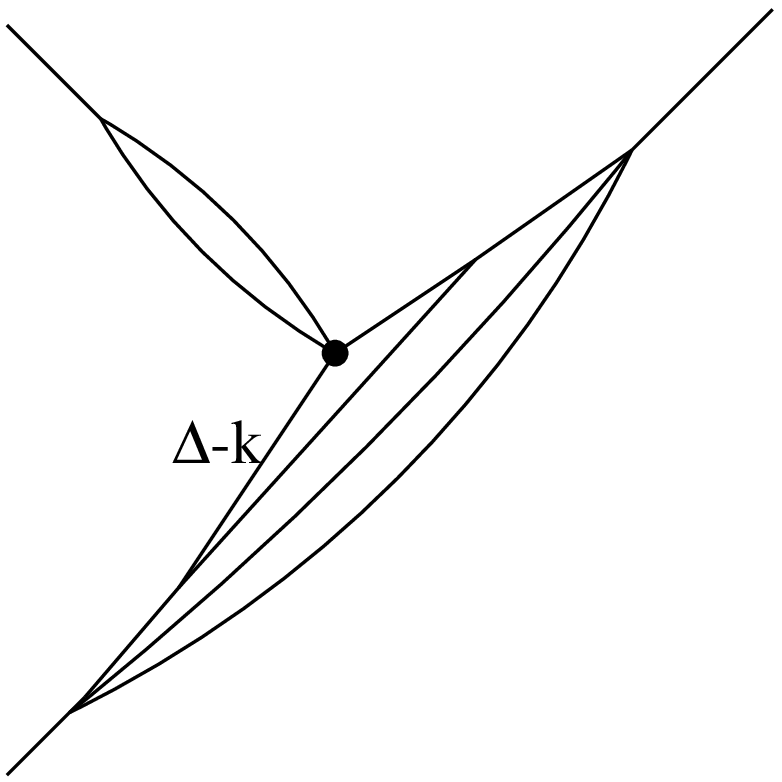}
           &
           \epsfxsize=0.28\hsize
           \epsfbox{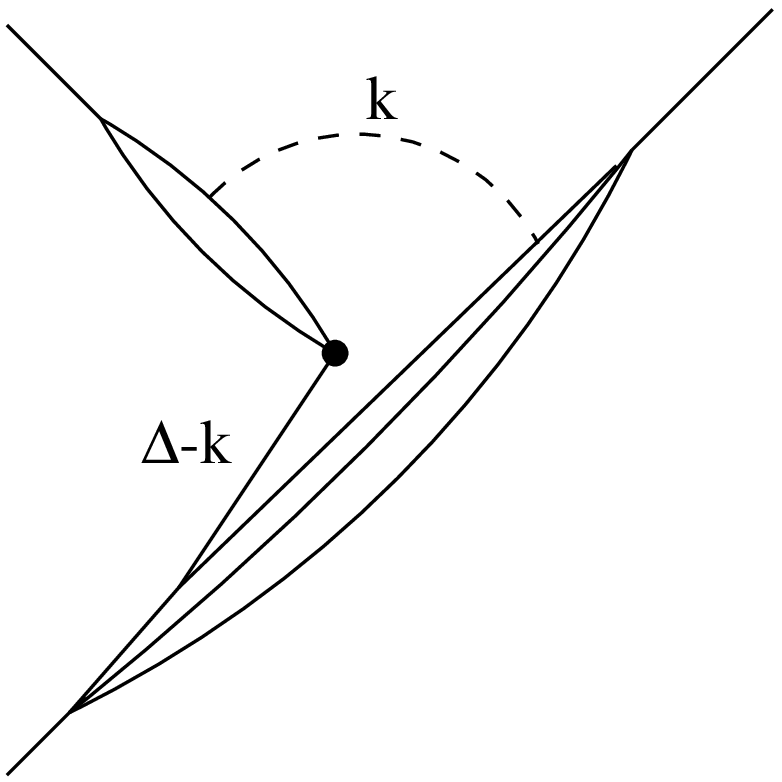}
        \\
           (c) & (d)
        \end{tabular}
    \end{center}
\caption{A low order graph for diffractive meson
         production, together with three of its reduced graphs.
         The solid lines are meant to be at a $45^{\circ }$ angle to
         represent light-like propagation, but have been
         separated to permit the structure to be seen.  }
\label{fig:Example.1}
\end{figure}

To understand what Fig.\ \ref{fig:Space.Time} means, let us look
at a few examples of regions of momentum space that correspond to
reduced graphs obtained from the Feynman graph of Fig.\
\ref{fig:Example.1}(a).  There the couplings between the quarks
and the hadrons may be considered as Bethe-Salpeter wave
functions. We will not give an exhaustive list of all possible
reduced graphs, but will only give some typical examples that
correspond to leading power contributions to the amplitude.

\subsubsection{First example}

We first consider a region defined as follows:  The upper two
loops have momenta
\begin{eqnarray}
   k &\sim& \left( x_{1}p^{+}, O(m^{2}/xp^{+}), O(m) \right) ,
\nonumber\\
   l &\sim& \left( O(m^{2}/Q), zQ^{2}/2xp^{+}, O(m) \right) .
\label{eq:Example.1.b}
\end{eqnarray}
where $m$ represents a typical hadronic scale.
We continue to use a coordinate system like the Breit frame where
$xp^{+} \sim Q$, and we label the components in the order
$(+,-,\perp )$.  The parameter $z$, for the $-$ component of $l$ lies
between 0 and 1, and is not close to its endpoints.  The
parameter $x_{1}$, for the $+$ component of $k$ is chosen such that
both $x_{1}$ and $x-x_{1}$ are of order $x$ and are both positive.
Finally, the region is such that all the lower three lines have
momentum components of size
$\left( O(p^{+}), O(m^{2}/p^{+}), O(m) \right)$.

Another way of defining the region is to say that the quark lines
$l$ and $V-l$ are collinear to $V$, the quark lines $q-l$ and
$V-l-k$ are hard, and all the remaining lines are collinear to
$p$.

This region forms a neighborhood of the configuration defined by
the reduced diagram in Fig.\ \ref{fig:Example.1}(b).  In this
configuration, the lines
of momenta $q-l$ and $V-l-k$ form the hard vertex, since they
have virtuality of order $Q^{2}$. The lower 3 quark lines, and the
two gluons have momenta proportional to $(p^{+},0,0_{\perp })$, while the
lines $l$ and $V-l$ have momenta proportional to $(0, Q^{2}/p^{+},0_{\perp })$.
In the reduced diagram, the light-like momenta are represented by
lines in approximately the $45^{\circ }$ directions that represent their
world lines.  Since both of $k$ and $\Delta -k$ have positive $+$
momenta, they are forward moving lines.

It is important to make a pedantically exact distinction between
the momentum configuration represented by the reduced graph and
the region of integration that we attach to it.  Confusion
between the two concepts results in misunderstanding of the
content of parton-model concepts.  The configuration contains a
collection of light-like momenta derived by certain rules, while
the region is a neighborhood of this configuration.
The graph, Fig.\ \ref{fig:Example.1}(a),
is not singular when the momenta become
light-like in the way labeled by the reduced graph.  Apart from
anything else, the external hadrons have fixed nonzero mass.  The
singularity arises when the masses are set to zero.
What the use of the reduced diagram terminology does is to
usefully identify a certain region of momentum space.

As we explained earlier, a singularity is obtained in the
massless case when one integrates over arbitrary scalings of the
coordinates of the vertices of a reduced graph.  So when we
actually need to integrate over a small neighborhood of the
momentum configuration, that corresponds in
coordinate space to integrating over scalings of the positions of
the vertices, but up to a large instead of an infinite limit;
larger scalings are exponentially suppressed. The space-time
diagram obtained from a reduced graph then gives a region for the
positions of the vertices of the Feynman graph where (some of)
the vertices are separated by much more than order $1/Q$ in the
Breit frame.

\subsubsection{Second example}

Our second reduced graph, Fig.\ \ref{fig:Example.1}(c), is the
same as the first, except that the parameter $x_{1}$, defining the
longitudinal momentum fraction of $k$, has the opposite sign.
The space-time direction of the line $k$ is therefore reversed.
Previously, in Fig.\ \ref{fig:Example.1}(b), we had a two-gluon
state emitted from the proton and then entering the hard
scattering; this corresponds to the idea of the Pomeron as a
particle-like object.  But now that we have reversed the direction
of $k$, we have a situation in which
one gluon out of the proton generates a hard scattering, by
scattering off the virtual photon, and then continues into final
state where it coherently recombines with the remnants of the
target, to form the diffracted proton.

\subsubsection{Third example}

Our final example is where the gluon $k$ has soft momentum: all
its momentum components in the Breit frame are much less than $Q$
in size.  This gives Fig.\ \ref{fig:Example.1}(d) for a reduced
graph.  Note that the quark of momentum $V-l-k$ is now collinear
to $V$ rather than being hard.  Physically we have a situation in
which most of the Pomeron momentum is carried by one gluon, and
the hard scattering is photon-gluon fusion.  The second, soft
gluon just transfers color.  This is the kinematic situation of
the super-hard or coherent Pomeron \cite{CFS}.

As we will see later, although such configurations do give
leading contributions to the amplitude from individual graphs,
there is a cancellation after summing over different graphs.  The
remaining leading configurations correspond only to the first two
reduced graphs (and a third similar graph with $x_{1}>0$ and
$x-x_{1}<0$).
Other configurations give sub-leading contributions for the case
of a longitudinally polarized photon.


\subsection{General construction of reduced graphs}
\label{sec:construct.reduced.graphs}

The simplicity of Figs.\ \ref{fig:Reduced.Graph} and
\ref{fig:Space.Time}, which represent
the most general situation for our process, follows simply from
momentum conservation applied to classical processes, as we will
now show.  Since we have taken a massless limit, all the
explicitly displayed lines are light-like or have zero momentum.
The diagram must lie entirely in a plane spanned by the $+$ and
$-$ axes.  If not, there is a reduced vertex with a maximum
transverse position relative to the main hard vertex. Transverse
momentum conservation cannot be satisfied at such a vertex since
we have no external lines with non-zero transverse momentum, in
the massless limit we are taking.

If one starts from some line with momentum in the $-$ direction
and follows it backward on a connected series of lines with $-$
momenta, one arrives at an earliest vertex.  This must be the
hard vertex $H$, where the virtual photon attaches, since this is
the only place where large $-$ momentum is injected into the
graph.  Then if we go forward again, we get to a latest vertex,
necessarily later than the hard scattering.  This is where the
outgoing meson attaches.  The fact that all the $B$ lines are
later than the hard scattering will be important later when we
analyze soft gluon attachments to the $B$ subgraph.

Similarly on the $A$ lines, if one goes back one arrives at
either $H$ or the incoming proton.  If one goes forward one
arrives at either $H$ on the outgoing proton.

\begin{figure}
    \begin{center}
        \begin{tabular}{c@{~~~~}c@{~~~~}c}
           \epsfxsize=0.2\hsize
           \epsfbox{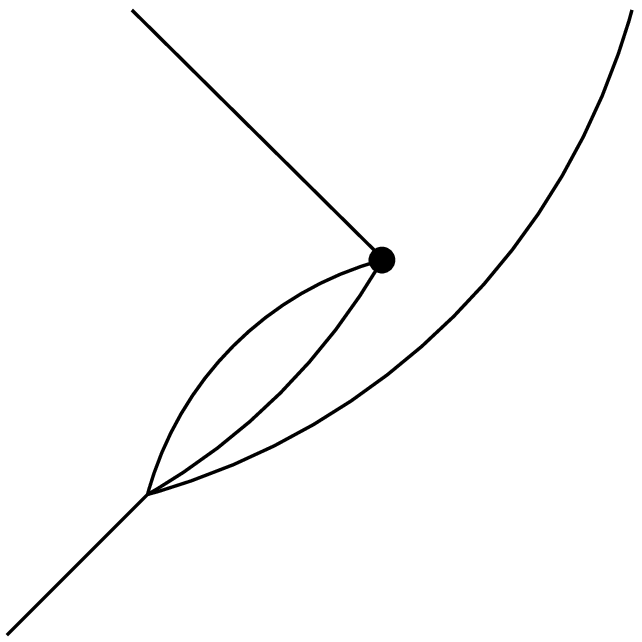}
           &
           \epsfxsize=0.2\hsize
           \epsfbox{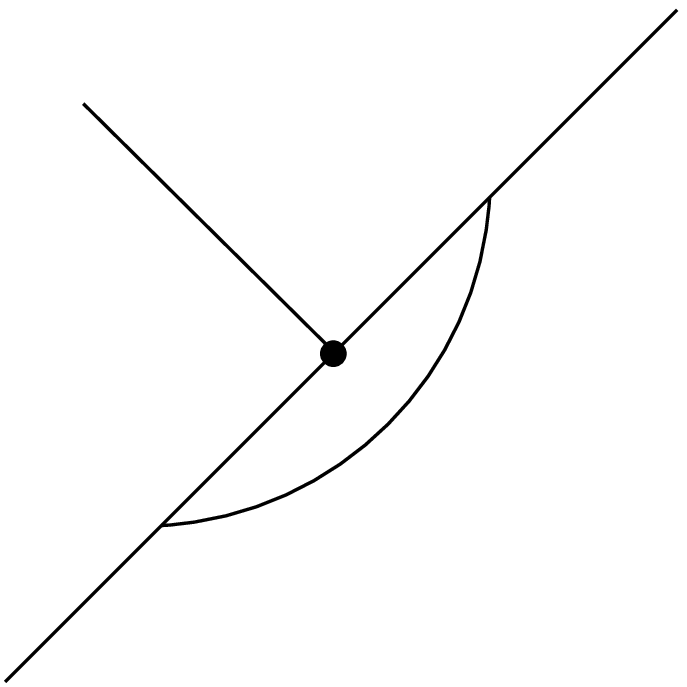}
           &
           \epsfxsize=0.15\hsize
           \epsfbox{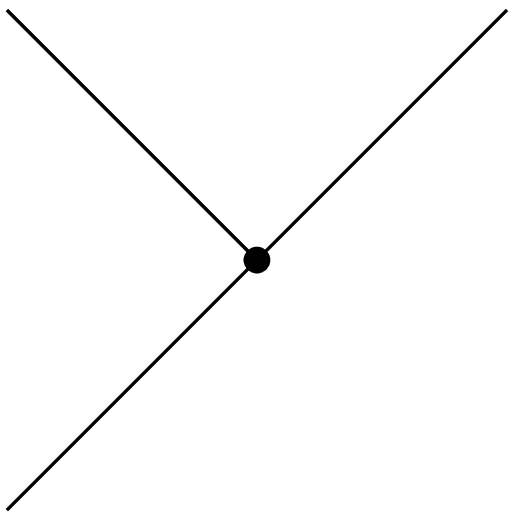}
        \\
           (a) & (b) & (c)
        \end{tabular}
    \end{center}
\caption{Examples of the three classes of space-time diagram.}
\label{fig:Space.Time.2}
\end{figure}

There are in fact three distinct topologies, as shown in Fig.\
\ref{fig:Space.Time.2}, where, to enable the topologies to be
visualized, we have slightly deformed the $A$ lines. In the first
class, the hard scattering has incoming $A$ lines, but no
outgoing $A$ lines. The partons that construct the outgoing
proton are all emitted before the hard scattering in this class
of graphs.

In the second class, the hard scattering has one or more outgoing
$A$ lines, so that the hard scattering directly influences the
outgoing proton.  But there are also $A$ lines that bypass the
hard scattering.

Finally, in the third class of graphs, no collinear lines bypass
the hard scattering.  In fact, such graphs have too many partons
entering the hard scattering to be leading; this will follow from
the power-counting arguments in the next section.

In all cases the number of lines entering and leaving the hard
subgraph is completely arbitrary.  It is the power-counting
properties  explained in Sec.\ \ref{sec:powers}  that will
restrict the situation for the leading power in $Q$; these are
results that follow from the specific dynamics of the theory.

Note that there will be quantum mechanical interference between
the different classes of graph, when one adds all the different
contributions to make the complete amplitude.  Moreover in each
reduced graph, the positions of the vertices along the lines must
be integrated over.  Thus the different space-time positions for
the vertices do not represent independent happenings.

We have constructed the reduced graphs and the space-time
diagrams with the ansatz of exactly massless external lines.
To avoid any confusion, let us reiterate that the actual process
has external hadron lines that are massive, even though these
masses are much less than $Q$.  The quarks also have nonzero
masses. The space-time method has enabled us to identify in
complete and simple generality the locations of the pinch
singular surfaces of corresponding massless Feynman graphs for
our process. The significance of these surfaces is that we will
classify contributions to the actual amplitude by neighborhoods
of these surfaces in loop-momentum space (with all the masses
preserved). Most importantly, the construction of the
factorization theorem will rely on identifying all significant
contributions to the amplitude with particular singular surfaces.


\section{Power counting}
\label{sec:powers}

We now wish to identify the power of $Q$ associated with each of
the pinch singular surfaces catalogued in the previous section,
and hence to identify those surfaces that give contributions to
the leading power. Again, the basic results are those of Libby
and Sterman.  Their results were mostly obtained in an axial
gauge, such as $A^{0}=0$ or $A^{3}=0$. However, the unphysical
singularities in the gluon propagator for a ``physical gauge''
prevent us from using certain contour deformation arguments, so
we prefer to work in a covariant gauge---compare Ref.\
\onlinecite{fact1}.  The method for obtaining the powers that we
present here is rather different to that given by Sterman
\cite{Sterman1}, and relies more on general properties of
dimensional analysis and Lorentz transformations than on a more
detailed analysis of the numbers of loops, lines and vertices of
graphs and subgraphs.

A complication to working in a covariant gauge is that graphs
with collinear gluons attached to the hard part are enhanced by a
power of up to $Q^{2}$ compared to the power obtained in axial
gauge.  The enhancement occurs when the gluons have scalar
polarization.  As Labastida and Sterman \cite{LabSt} showed,
Slavnov-Taylor identities can be used to show a cancellation of
the enhanced contributions, so that the final result for the
power counting is the same as in axial gauge.  We will use a
somewhat different, but equivalent, method of obtaining this
result, in Sec.\ \ref{sec:ST}.

The result we will prove in the remainder of the present section
is that, {\em before these cancellations}, the power of $Q$
associated with a pinch singular surface $\pi $ is $Q^{p(\pi )}$, with
\begin{eqnarray}
   p(\pi ) &=& 3 - n(H)
            - {\rm No.}(\mbox{quarks from soft to collinear})
            - 3 {\rm No.}(\mbox{quarks from soft to hard})
\nonumber\\
        &&
            - 2 {\rm No.}(\mbox{gluons from soft to hard}) .
\label{power}
\end{eqnarray}
Here $n(H)$ is the number of external collinear quark and
transversely polarized gluon lines of the hard subgraph. The
other two terms involve the number of quark lines that attach the
soft subgraph to either of the collinear subgraphs and the number
of lines going from the soft subgraph to the hard subgraph.

Notice that the power decreases as the number of external lines
of the hard scattering increases;
this is the essential rationale for the parton model,
where the minimum number of partons is used in the hard
scattering.  For the gluons we must, as we will see, split their
polarizations into what we will call ``scalar'' and
``transverse'' components.  There is only a suppression for
extra transverse gluons entering the hard scattering;
any number of collinear gluons with ``scalar polarization'' can
attach to the hard subgraph, without a penalty in powers of $Q$.

Our arguments will use rather general properties of dimensional
analysis and Lorentz boosts. When we examine the dependence on
the polarization of the virtual photon, in Sec.\
\ref{sec:transverse}, we will find that the power given in
Eq.~(\ref{power}) is normally obtained only for one photon
polarization, longitudinal or transverse, depending on the
region.


\subsection{Proof of power counting formula,
            Eq.\ (\protect\ref{power})}
\label{sec:powers.proof}

The strategy of our proof is first to prove it for certain
particularly simple cases, and then to extend it.

\subsubsection{Case of collinear and hard subgraphs only}

First consider a case of Fig.\ \ref{fig:Reduced.Graph} when we only
have collinear and hard subgraphs, but no soft subgraph.  Let
the hard subgraph $H$ have $N_{q}$ external quark (and antiquark)
lines and $N_{g}$ external gluons, as well as a single photon line.

By definition, all components of loop momenta in the hard
subgraph have size $Q$, in the Breit frame, and all the lines in
the subgraph have virtuality of order $Q^{2}$. Since the hard
subgraph has dimension $d_{H}=3-\frac {3}{2}N_{q}-N_{g}$ and all the
couplings are dimensionless, it contributes a power
\begin{equation}
   Q^{d_{H}} = Q^{3-\frac {3}{2}N_{q}-N_{g}}
\label{first.power}
\end{equation}
to the amplitude.

For the momenta collinear to the meson we assign orders of
magnitude
\begin{equation}
   \mbox{typical $V$ momentum} \sim
   \left(xp^{+}\frac {m^{2}}{Q^{2}}, \frac {Q^{2}}{xp^{+}}, m \right)
   \sim
   \left(\frac {m^{2}}{Q}, Q, m \right),
\end{equation}
in $(+,-,\perp )$ coordinates, with $m$ being a typical hadronic mass.
Similarly we assign momenta collinear to the proton a magnitude
\begin{equation}
   \mbox{typical $A$ momentum} \sim
   \left(Q, \frac {m^{2}}{Q}, m \right).
\end{equation}
Since the Bjorken variable $x$ is small, there are also collinear
momenta with $+$ components much larger than $Q$.  We will deal
with this complication later; for the moment let us treat the
case that $x$ is not small.

The collinear configurations can be obtained by boosts from a
frame in which all components of all momenta are of order $m$.
Since virtualities and the sizes of regions of momentum
integration are boost invariant, we start by assigning the
collinear subgraphs an order of magnitude $m^{{\rm dimension}}$,
which contributes exactly unity to the power of $Q$.  This also
enables us to see that non-perturbative effects, as coded in a
Bethe-Salpeter wave function, for example, do not change the
power of $Q$.  Note that we define the collinear factors to
include the integrals over the momenta of the loops that couple
the collinear subgraphs and the hard subgraph.

Next we must allow for the fact that the collinear subgraphs are
coupled to the hard subgraph by Dirac and Lorentz indices.  Now,
the effect of boosting a Dirac spinor from rest to a large energy
$Q$ is to make its largest component of order $(Q/m)^{1/2}$ bigger
than the rest frame value, and the effect on a Lorentz vector is
to give similar factor $(Q/m)^{1}$.  The exponents $1/2$ and 1 are
just the spins of the fields.  The resulting powers multiply
Eq.\ (\ref{first.power}) to give
\begin{equation}
   Q^{3-N_{q}} .
\label{second.power}
\end{equation}
This agrees with Eq.~(\ref{power}) in the case that all the
external lines of the hard subgraph are quarks, but is a factor
$Q^{N_{g}}$ larger whenever there are external gluons.

Later, in Sec.\ \ref{sec:ST}, we will show how cancellations
between different graphs cause a suppression of the highest
powers associated with collinear gluons attaching to the hard
subgraph.  As we have already stated, these are contributions
from gluons of scalar polarization.  For the moment we just need
to define the concepts of scalar and transverse polarization in
the sense that we will use, and to show how this affects the
power counting.

Consider the attachment of one gluon, of momentum $k^{\mu }$, from the
$A$ subgraph to the hard subgraph.  We have a factor
${\cal A}^{\mu }(k)  \, g_{\mu \nu } \, {\cal H}^{\nu }(k)$,
where ${\cal A}^{\mu }$ and ${\cal H}^{\nu }$ denote the $A$ and $H$
subgraphs, and $g_{\mu \nu }$ is the numerator of the gluon propagator in
Feynman gauge.\footnote{
    A change to another covariant gauge merely results in
    notational complication.
}
We decompose this factor into components:
\begin{equation}
    {\cal A}\cdot {\cal H} =
    {\cal A}^{+}{\cal H}^{-} + {\cal A}^{-}{\cal H}^{+}
    - {\cal A}_{\perp }\cdot {\cal H}_{\perp } ,
\label{AH.connection}
\end{equation}
and we observe that after the boost from the proton rest frame,
the largest component of ${\cal A}^{\mu }$ is the $+$ component.  The
largest term is therefore ${\cal A}^{+}{\cal H}^{-}$, and this is the
term that gives the power stated above, in Eq.\
(\ref{second.power}).  The other two terms are suppressed by one
or two powers of $Q$.

So we now define the following decomposition:
\begin{equation}
   {\cal A}^{\mu } = k^{\mu } \frac {{\cal A}^{+}}{k^{+}}
        \, + \,
        \left( {\cal A}^{\mu }
               - k^{\mu } \frac {{\cal A}^{+}}{k^{+}}
        \right) .
\label{gluon.pol.1}
\end{equation}
The first term we call the scalar component of the gluon: it
gives a polarization vector proportional to the momentum of the
gluon.  The second term, the transverse part of the gluon,
has a zero $+$ component: it therefore gives a contribution to
${\cal A}\cdot {\cal H}$ that is one power of $Q$ smaller than the
contribution of the scalar component.  The $k^{\mu }$ factor in the
scalar term multiplies the hard subgraph, and this gives a
quantity that can be simplified by the use of Ward identities, as
we will find in Sec.\ \ref{sec:ST}.  .

We now apply this decomposition to every gluon joining the
subgraphs $A$ and $H$, and the analogous decomposition for gluons
joining $B$ and $H$.  The contribution of our region to the
amplitude is now a sum of terms in which each of these gluons is
either scalar or transverse.  Each term has a power
\begin{equation}
   Q^{3-N_{q}-N_{g}} Q^{N_{s}} = Q^{3-N_{q}-N_{T}}  ,
\label{basic.power}
\end{equation}
where $N_{s}$ is the number of scalar gluons, and $N_{T}=N_{g}-N_{s}$ is the
number of transverse gluons that enter the hard scattering.
This is the exact power that we wrote in Eq.~(\ref{power}), given
that we have no soft subgraph.

In should be noted that in Sec.\ \ref{sec:ST} we will slightly
modify the definitions of ``scalar'' and ``transverse''
polarizations---see Eq.\ (\ref{gluon.pol.2}) below.  This will be
to take account of the Taylor expansion we will apply to the hard
subgraph, and also to apply an exactly analogous argument to the
couplings of soft gluons to a collinear subgraph.

We also will need to to pick out the largest component of the
Dirac structure of the collinear subgraphs, but do not need to
make the operation explicit here, since we do not have a
cancellation of the highest power.  We just note that the
projection of the largest Dirac component is directly reflected
in the factors of $\gamma ^{+}$ and $\gamma ^{-}$ in the
definitions of the quark
distribution and wave function, Eqs.\ (\ref{pdf.q.def}) and
(\ref{wf.def}).

\subsubsection{Small $x$}

The derivation of the power Eq.~(\ref{basic.power}) assumed that
$x$ was not small.  Now if $x$ is made small, we must boost some
parts of the collinear-to-$A$ subgraph to get $+$ momenta of
order $p^{+}$ instead of $xp^{+}$, so that groups of lines have very
different rapidities.  It is known that in Feynman graphs the
effect is simply to give a factor $1/x$ (times logarithms), but
only provided that all the lines exchanged between the regions of
different rapidity are gluons.  For example, see Ref.\ \onlinecite{BFKL}.
If any quarks are exchanged, there is a suppression by a factor
of $x$.  None of this affects the power of $Q$.

\subsubsection{Soft lines}

We now add in a soft subgraph $S$.  A problem is to choose an
appropriate scaling of the momenta, a problem that has not
entirely been solved satisfactorily
in the literature.  One possibility is to assign all components
of soft momenta a size $m$.  This has the advantage of being
immune to non-perturbative effects in the soft subgraph, and the
disadvantage of sending at least some lines in the collinear
subgraphs off-shell, by order $Qm$.\footnote{
    ``Disadvantage'' here means a disadvantage from the point of
    view of a simple construction of a power-counting formula.
}
A second possibility is to assign all
the soft momenta a size $m^{2}/Q$.  This avoids sending collinear
lines far off-shell, but forces us to treat a region where the
momenta are unphysically soft in a confining theory, and where
the power counting is sensitive to mass effects.

In fact we will choose the second scaling.  All other
possibilities will be covered by the arguments in Sec.\
\ref{other.scalings}.

A more general treatment \cite{Sterman1} would assign a size
$k \sim \lambda Q$ to the components of a soft momentum.  Here $\lambda $ is an
integration variable that is much less than one.  To determine
the power of $Q$, one has to determine how small $\lambda $ can be made:
there are significant changes when $\lambda =O(m/Q)$ and when
$\lambda =O(m^{2}/Q^{2})$, from mass effects in the soft propagators and the
collinear propagators respectively.

Given that we assign all momenta in $S$ a magnitude $m^{2}/Q$ for
all their components in the Breit frame, the basic power for the
soft subgraph is $m^{2}/Q$ to a power which is the dimension of the
soft subgraph.  This power includes the integration over the soft
loop momenta that circulate between $S$ and the rest of the
graph, and it assumes that we can neglect masses in the
propagators.  The numerical value of the power is
\begin{equation}
   -N_{gS} - \frac {3}{2} N_{qS} ,
\end{equation}
where $N_{gS}$ and $N_{qS}$ are the numbers of external gluons and
quarks of the soft subgraph $S$.

These external lines go into either the hard subgraph or one of
the collinear subgraphs.  The dimension of the hard subgraph is
reduced by $3/2$ for each extra soft gluon that enters it and $1$
for each quark.  The dimensions of the collinear subgraphs are
changed, but this does not affect the power of $Q$.  But there
are spinor and vector indices joining the soft and collinear
subgraphs, and just as with the collinear-to-hard connections we
get a factor of $Q^{1/2}$ for each quark and a factor $Q$ for each
gluon.

Putting all the factors together gives Eq.~(\ref{power}) for the
power of $Q$ for the contribution of our region to the amplitude.
The qualitative features to note are that:
\begin{itemize}

\item Extra external lines for the hard subgraph always reduce
   the power of $Q$, except in the case of scalar gluons.

\item There is no suppression for soft gluons attaching to the
   collinear subgraphs, as is well-known.

\item There is a penalty for soft quarks attaching to the
   collinear subgraphs, as is also well-known.

\end{itemize}
But observe that there is no penalty for having quark loops
{\em inside} the soft subgraph.  This is a fact that is sometimes
forgotten, because in the corresponding infra-red-divergence
problem in QED, no loops of massive fermions need to be
considered.  When we allow a general scaling $\lambda Q$ for soft
momenta there is no necessary suppression of quark loops inside
the soft subgraph.

\subsubsection{Other scalings}
\label{other.scalings}

Any other scalings of the momenta can be considered as
intermediate between those we have listed.  The one exception we
will discuss in a moment.  We have catalogued all pinch
singular surfaces of massless graphs for our process and have
defined the regions as neighborhoods of these surfaces.  The
scalings of momenta defined above may be called canonical
scalings for each of the regions.

When the asymptotics of graphs are treated, all other scalings
can be treated as a way of interpolating between the canonical
scalings for different regions.  The methods we use will treat
the intermediate regions correctly once the canonical scalings
are taken into account, and intermediate scalings between two or
more different leading regions will be responsible for the
omnipresent logarithms in the asymptotics of Feynman graphs.

The one exception to the above rule are the truly infra-red
regions, where some momenta go to zero.  In a theory of confined
quarks and gluons these regions are not genuinely physical, but
they do appear in Feynman graphs.  They are treated by a
sufficiently careful treatment of the soft region as we have
defined it.


\subsection{Catalog of leading regions}
\label{sec:catalog}

\begin{figure}
    \begin{center}
        \begin{tabular}{c@{~~~}c@{~~~}c}
           \epsfxsize=0.28\hsize
           \epsfbox{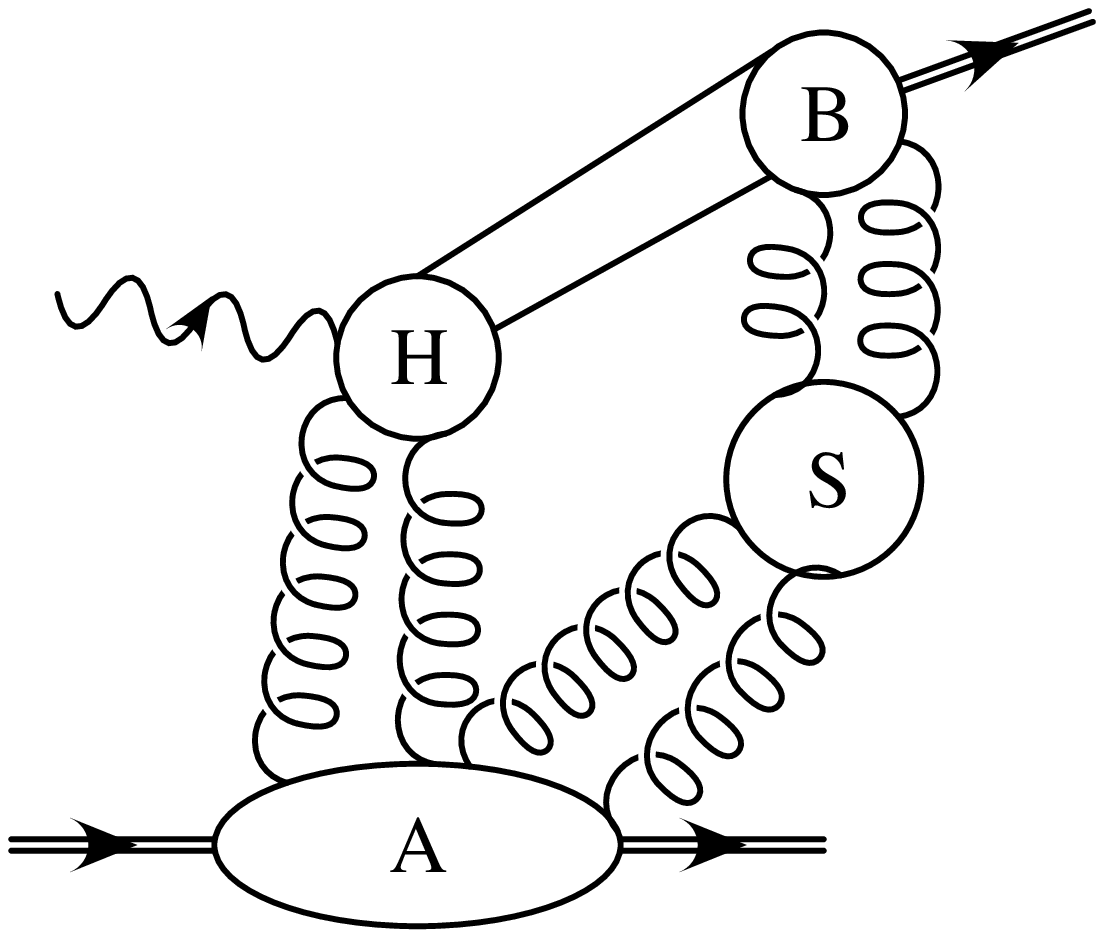}
           &
           \epsfxsize=0.28\hsize
           \epsfbox{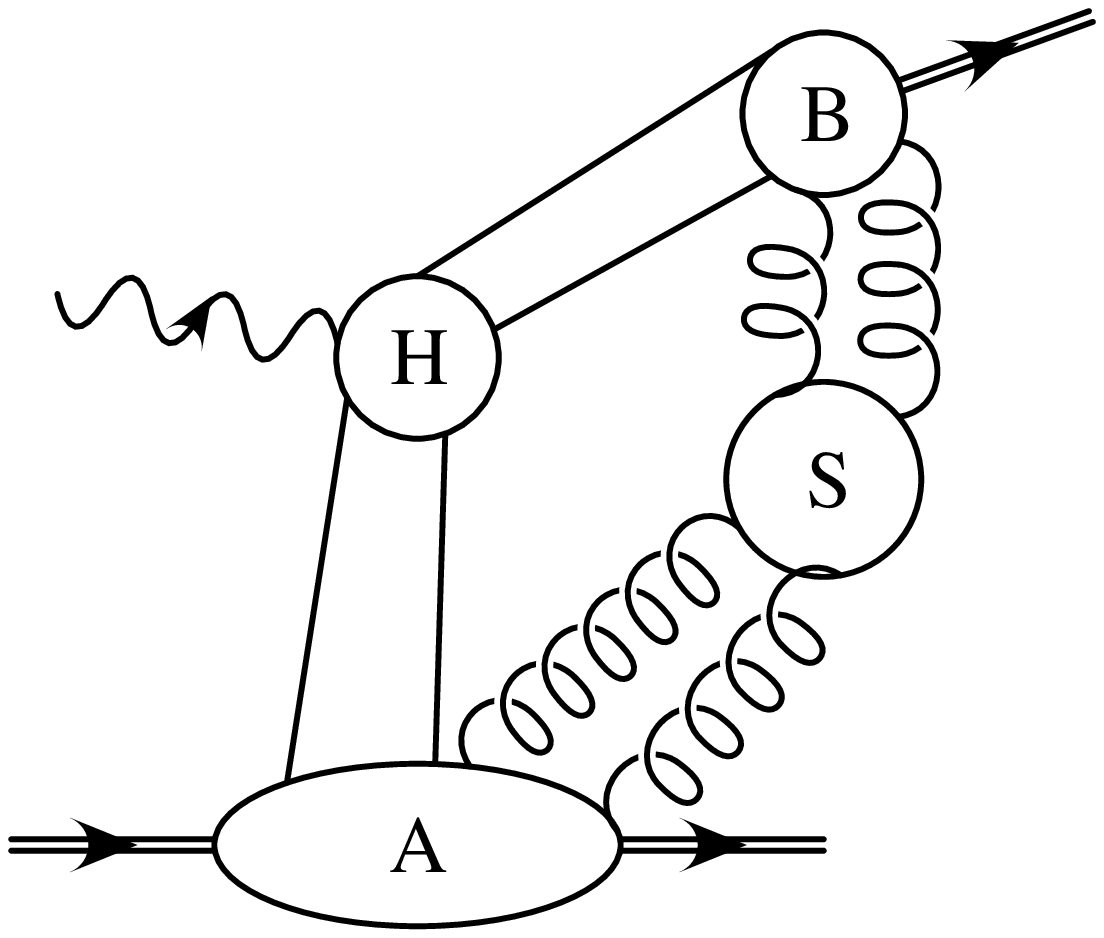}
           &
           \epsfxsize=0.28\hsize
           \epsfbox{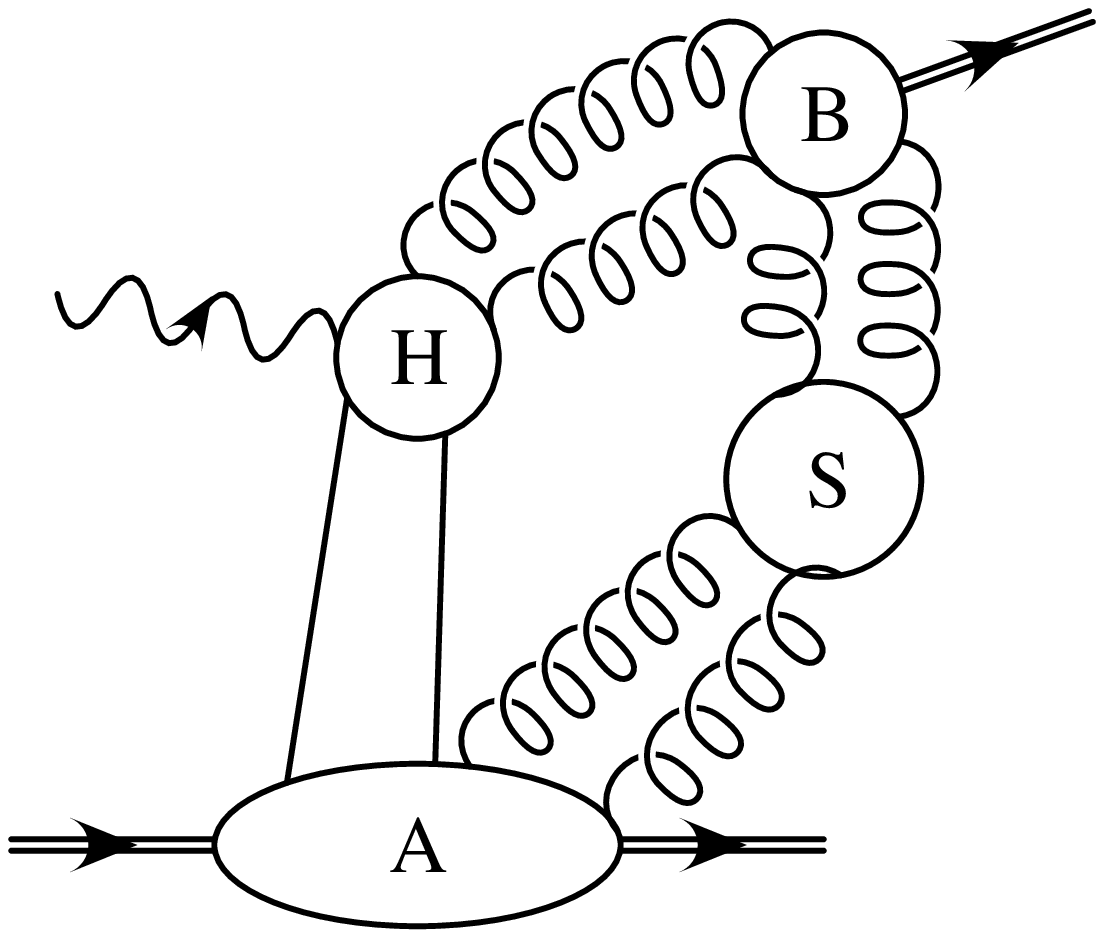}
        \\
           (a) & (b) & (c)
        \\
           \epsfxsize=0.28\hsize
           \epsfbox{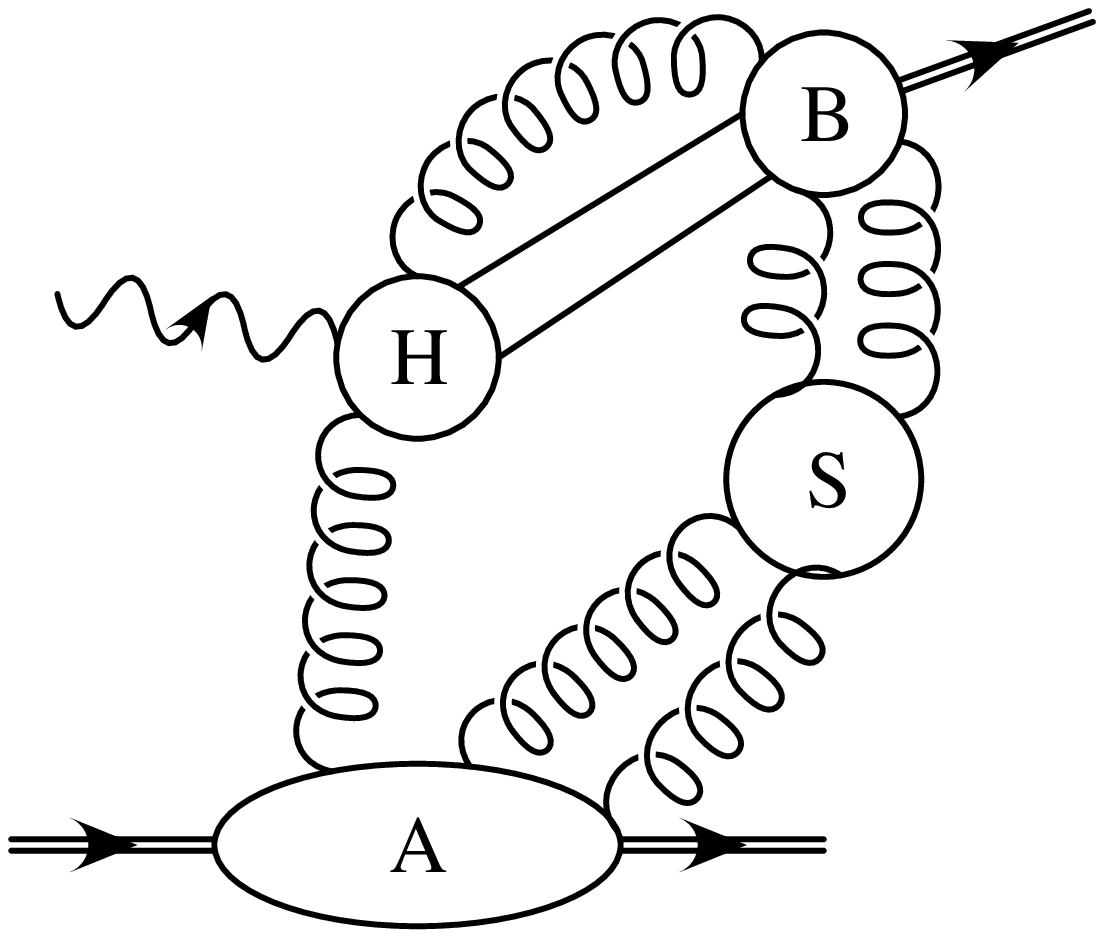}
           &
           \epsfxsize=0.28\hsize
           \epsfbox{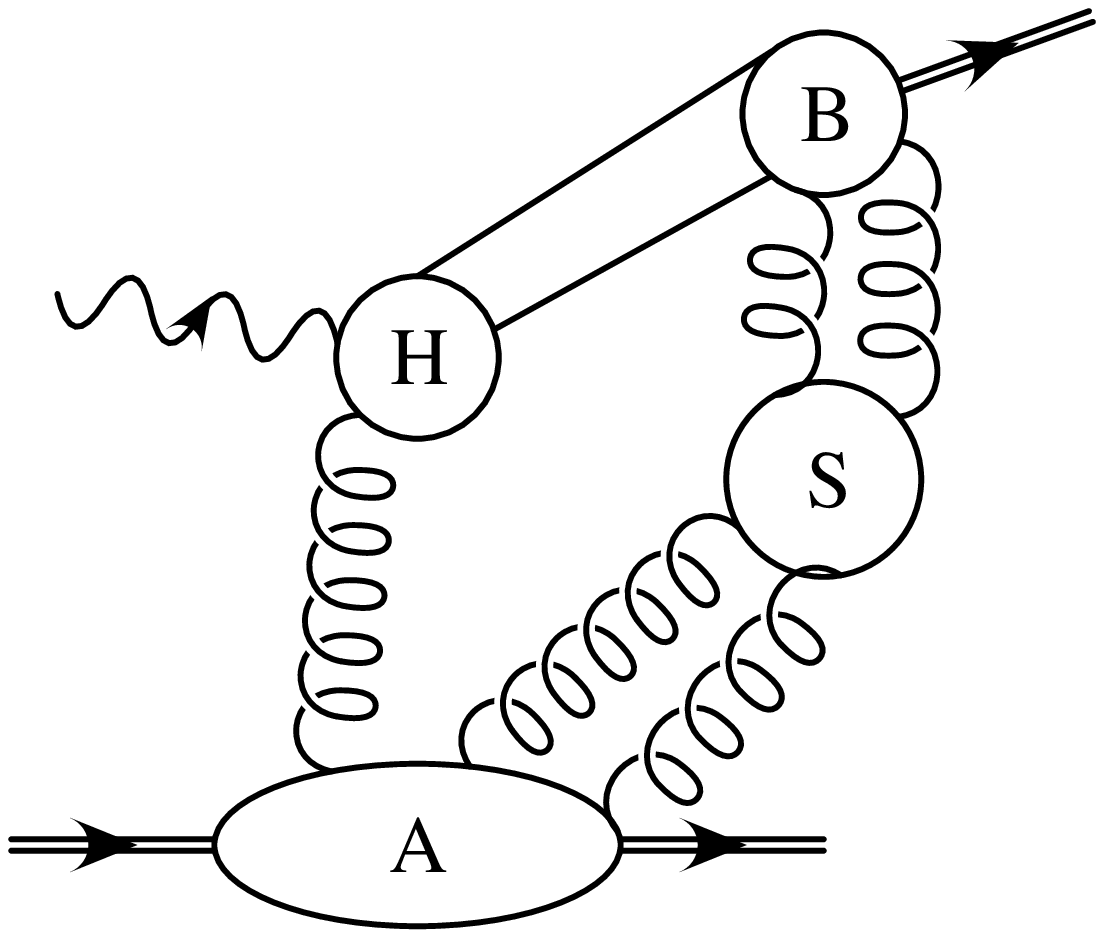}
           &
           \epsfxsize=0.28\hsize
           \epsfbox{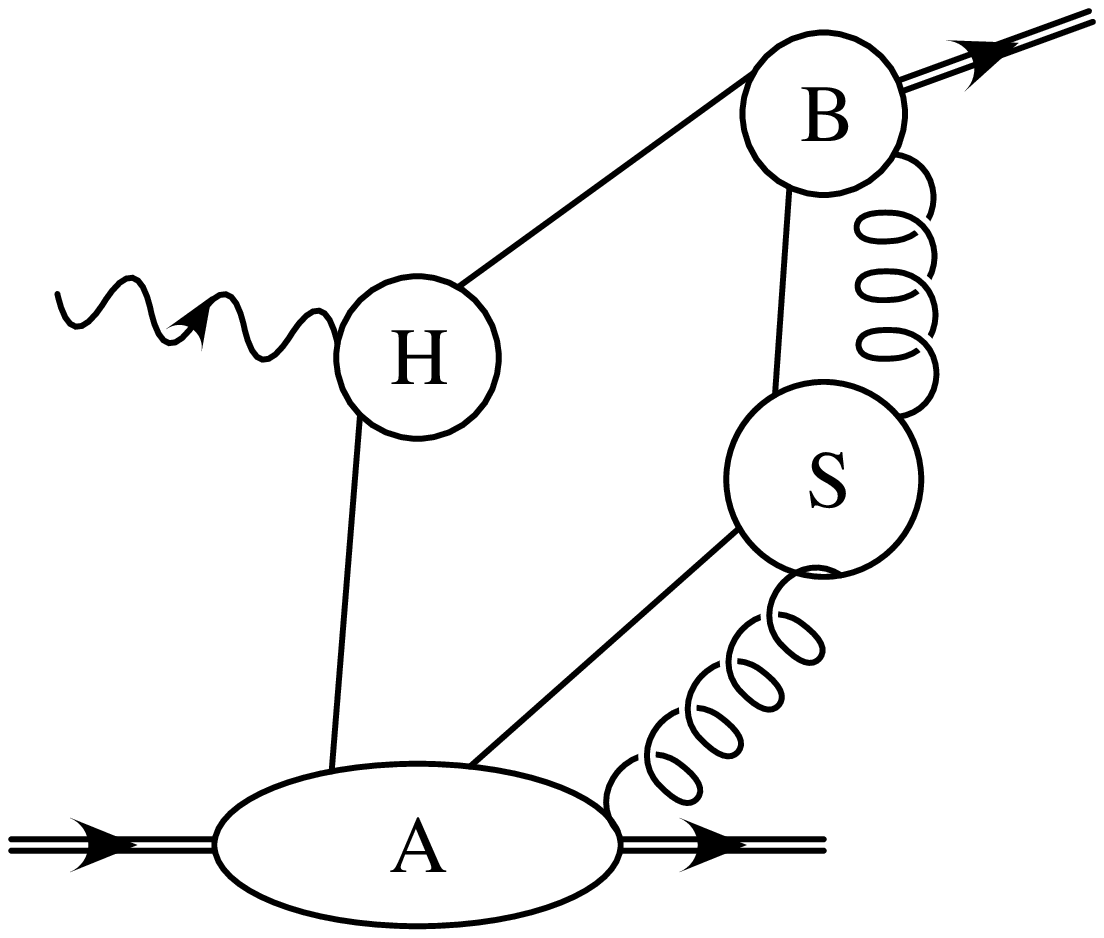}
        \\
           (d) & (e) & (f)
        \end{tabular}
    \end{center}
\caption{Leading regions for our process, when gluons entering
         the hard subgraph are transversely polarized, and when
         we do not explicitly indicate terms where the meson
         couples to a set of gluons.  Each soft subgraph may have
         any number of external gluons, including zero. }
\label{fig:Leading.Regions}
\end{figure}

When all cancellations have been taken into account, we will find
that the
amplitude behaves like $1/Q$ (times logarithms), for large $Q$.
In addition, for the $x\to 0$
asymptotics, there is a power $1/x$ that corresponds to spin-$1$
exchange in the $t$-channel (from the simplest models of the
Pomeron).  Thus the overall power is $s/Q^{3}$, so that the cross
section $d\sigma /dt \sim 1/Q^{6}$, in agreement with the results of
\onlinecite{BFGMS}.
Our actual proof of the factorization theorem will be rather
indirect, to take account of the cancellations caused by gauge
invariance.\footnote{
   Note that before the cancellations, the highest power
   possible, according to Eq.~(\ref{power}), is $Q^{3}$, when all
   the external lines of the hard and soft subgraphs are
   gluons of scalar polarization.  This situation is actually
   prohibited by our choice of quantum numbers for the meson, and
   the actual highest power is $Q^{1}$, from the region in Fig.\
   \ref{fig:Leading.Regions}(a).  Cancellations are needed to get
   a final power of $1/Q$.
}
But it is useful to identify the regions that give
the $1/Q$ behavior or larger; no other regions can give a
contribution to the leading power.

Compared to the usual factorization theorem for inclusive
scattering, the discussion is more involved, since we need to
treat cases where the hard scattering amplitude has four external
lines, instead of just two.
So, to simplify the discussion, we will restrict our attention to
the case that the collinear gluons attaching to the hard subgraph
have transverse polarization.  The other cases will be taken care
of by gauge-invariance.  The resulting list of regions is shown
in Fig.\ \ref{fig:Leading.Regions}.

First we observe that, by Eq.~(\ref{power}), we need to consider
only hard subgraphs with at most four external quark and
transverse gluon lines.

Two cases with four external lines for $H$ are (a) and (b), which
have a quark-antiquark pair going from the hard scattering to the
meson, and with either a gluon pair or a quark-antiquark pair
joining the hard part to the proton.  There is a possible soft
part joined to the collinear subgraphs by arbitrarily many
gluons. These terms correspond to the final factorization
theorem, after a cancellation of the effects of the soft gluons.
A third possibility is where all the collinear-$B$ lines of
the hard subgraph are transverse gluons, as in graph (c).
In this case we can
make a cut of the graph such that the meson couples to gluons;
such graphs we will call ``glue-ball'' graphs.  We will find that
they all cancel at the leading power $1/Q$.
A fourth possibility is in graph (d), where one collinear gluon
comes from the proton, and three collinear partons go to the
meson.  In the final factorization theorem, this would need a
color octet operator in the proton factor, and such an operator
has a zero matrix element between proton states.

Next we can have graphs with two or three external lines for the
hard scattering, Fig.\ \ref{fig:Leading.Regions}(e) and (f).
These are harder to treat, as we will see in Sec.\
\ref{sec:endpoint}.  In fact, graph (f)
will make a leading contribution in the case of a transversely
polarized photon and will not make a factorization theorem of the
form Eq.~(\ref{factorization}).  Graph (e) will combine with
those of type (a), with two collinear gluons entering the hard
part, when we construct the appropriate operator product
expansion.

According to Eq.~(\ref{power}), we have a leading $1/Q$
contribution from graph (f), where the hard part has two external
quark lines, and the quark loop is completed in the soft part.
But now observe that the hard part, to the leading power of $Q$,
is the on-shell electromagnetic form factor of a massless quark.
(Subtractions to prevent the double counting of different regions
will remove the infra-red divergences of the form factor.)  This
form factor is proportional to
\begin{equation}
   \epsilon _{\gamma ^{*}}^{\mu } \bar u_{B} \gamma _{\mu } u_{A},
\label{QuarkFF}
\end{equation}
where $u_{A}$ and $u_{B}$ are Dirac wave functions for the external
quarks of the hard scattering, and
$\epsilon _{\gamma ^{*}}$ is the polarization vector of the
virtual photon.
By the rules for computing a hard
scattering amplitude, the momenta of the quarks are massless and
are in the $+$ and $-$ directions.
Since we have chosen the photon to be longitudinally polarized,
$\epsilon _{\gamma ^{*}}^{\mu }$ is a
linear combination of the momenta of the two quarks.  Hence the
Dirac equation for massless spinors gives us zero for
Eq.~(\ref{QuarkFF}).

Notice that this argument does not apply when the photon is
transverse; graph (f) exactly corresponds to the endpoint
contribution discussed in Ref.\ \onlinecite{BFGMS}.  We will discuss
this issue in more detail in Sec.\ \ref{sec:transverse}.


\subsection{Other gluons joining the collinear subgraphs to the
            hard part}

\begin{figure}
    \begin{center}
        \leavevmode
        \epsfxsize=0.4\hsize
        \epsfbox{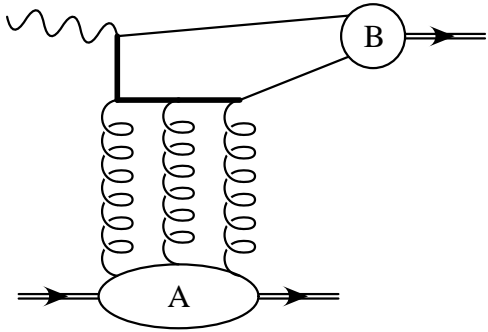}
    \end{center}
\caption{Example of region with extra gluons joining collinear
   subgraphs to hard subgraph.  All lines are supposed to be
   collinear to the meson or the proton, as appropriate, except
   for the two thick lines, which have virtuality of order $Q^{2}$.}
\label{fig:Extra.gluons}
\end{figure}

We have now seen that all the leading regions, that give the
power $1/xQ$ for the amplitude have the form of Fig.\
\ref{fig:Leading.Regions} (a), (b) and (e), given that the photon is
longitudinally polarized.  For clarity, the figures are not drawn
quite correctly since we have not yet treated the
cancellation of gluons with scalar polarization.  In the graphs,
any number of extra gluons may join each collinear subgraph to
the hard subgraph.  An example is shown in Fig.\
\ref{fig:Extra.gluons}.  As shown in Sec.\
\ref{sec:powers.proof}, the addition of extra scalar gluons does
not change the power of $Q$.

The fact that scalar gluons have a polarization proportional to
their momentum suggests that they can be eliminated by a gauge
transformation.  In fact, we will use gauge invariance, in Sec.\
\ref{sec:ST}, to show that only matrix elements of
gauge-invariant operators are needed in the definitions of the
parton-density and the wave-function factors in the factorization
theorem, Eq.~(\ref{factorization}). The result will be that the
contributions of scalar gluons will give the
path-ordered exponentials in the gauge-invariant operators that
define the distribution and density functions in
Eqs.~(\ref{pdf.q.def}) and (\ref{wf.def}).

In an appropriate axial gauge, the contributions of the scalar
gluons are power-suppressed, and correspondingly the path-ordered
exponentials in the operators are unimportant. This fact would
render the use of an axial gauge very attractive in proving
factorization, were it not for the complications in treating soft
gluons that result from the unphysical poles of the gluon
propagator in these `physical gauges'.  Compare the work in
Refs.\ \onlinecite{fact1,ColSt} on proofs of factorization theorems for
inclusive processes.


\section{Subtractions}
\label{sec:subtractions}

For each graph $\Gamma $, there may be several different regions of
loop-momentum space that contribute to the leading power. Each
region is associated with a pinch-singular surface $\pi $ of the
corresponding massless graph, and we write the graph as a sum
of contributions each associated with one surface:
\begin{equation}
   {\rm Asy} \, \Gamma  = \sum _{\pi } \Gamma _{\pi } ,
\end{equation}
where `Asy' denotes the asymptotic behavior of the graph.
In this section we will summarize the construction of the terms
on the right-hand-side of this equation.

Roughly speaking,
the term $\Gamma _{\pi }$ is obtained by Taylor expanding the hard and
collinear subgraphs in powers of the small variables, an
operation we denote by $T_{\pi }$.  Since there may be more than one
region contributing for a given graph, we must make subtractions
which will avoid double counting; the operation of applying the
subtractions we will denote by $R$, since it is a kind of
renormalization.  Thus we will write
\begin{equation}
   {\rm Asy} \, \Gamma  = \sum _{\pi } R T_{\pi }(\Gamma ) .
\label{asy.basic}
\end{equation}

This structure is completely analogous to that of the Bogoliubov
$R$-operation for renormalization.  The most convenient way we
have found for formulating the procedure is due to Tkachov and
collaborators \cite{Tkachov}. Although the detailed exposition of
the method given in \onlinecite{Tkachov} is tied to Euclidean problems,
the general principles are not.\footnote{
   The problems explained by Collins and Tkachov\cite{CT} concern
   the question of the use of dimensional regularization to
   define certain integrals and most certainly do not impinge on
   the general principles.
}
In this method, the integrand of each graph $\Gamma $ as a
distribution.  Thus we define
\begin{equation}
   \langle \Gamma ,f\rangle  = \int dk \, \Gamma (k,p) \, f(k),
\label{distrib}
\end{equation}
where $k$ denotes the collection of loop momenta, $p$ the
external momenta, and $f(k)$ is a test function.  The
contribution of the graph to the scattering amplitude is given by
replacing the test function by unity.

The advantages of using these distributional techniques
\cite{Tkachov} stem from the fine control they give by enabling
us to treat different regions of momentum space separately
without having to make sharp boundaries between the different
regions.  This last point is particularly important in problems
like ours, where it is important to be able to deform contours of
integration away from non-pinch singularities; the use of
sharp boundaries between regions prevents the use of contour
deformation.

In this language, the contribution $\Gamma _{\pi }$
to ${\rm Asy} \, \Gamma $ from the
neighborhood of a pinch-singular surface $\pi $ is localized on the
surface; that is, it is proportional to a $\delta $-function (with
possible derivatives) that restricts the integration to the
surface. To obtain a convenient form for $\Gamma _{\pi }$, we observe that
the graph is a product of a factor that is singular on $\pi $ and a
factor that is non-singular there.  Thus we write
\begin{equation}
   {\rm Asy} \, \Gamma (k) =
   \sum _{\pi } C_{\pi }(k;p,\mu ) E_{\pi }(\mu ) ,
\label{asy}
\end{equation}
where $C_{\pi }(k)$ is a distribution that is localized on the
surface $\pi $ and is obtained by expanding the hard subgraph $H$ in
Fig.\ \ref{fig:Reduced.Graph} in powers of its small external
variables (with appropriate subtractions).  The
quantity $E_{\pi }$ corresponds to the product of the singular factors,
$A$, $B$, and $S$ in the reduced graph. Essentially, $C_{\pi }$
corresponds to the short distance factor on the surface $\pi $, and
$E_{\pi }$ to the long-distance factor.

We will only present a summary of proof that this all works.
An important observation is that
the issues are identical to those for other kinds of
factorization.  We first define a hierarchy of regions, by simple
set-theoretic inclusion: i.e., we define $\pi _{1} > \pi _{2}$ to mean that
the pinch singular surface $\pi _{1}$ contains the pinch singular
surface $\pi _{2}$. For any given pinch-singular surface $\pi $, we
construct its corresponding term in Eq.~(\ref{asy}) on the
assumption that the terms for all bigger regions have already
been constructed. Thus the construction of Eq.~(\ref{asy}) is
recursive, starting from the largest region.

\begin{figure}
    \begin{center}
        \leavevmode
        \epsfxsize=0.25\hsize
        \epsfbox{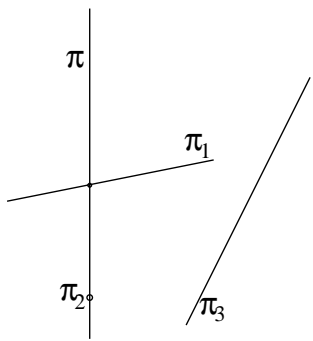}
    \end{center}
\caption{Illustrating the three classes for the ``other terms''
   in Eq.\ (\protect\ref{recursion}).}
\label{fig:other.terms}
\end{figure}

Suppose, then, that we have constructed the terms $\Gamma _{\pi '}$ for all
regions bigger than $\pi $.  Let us decompose
${\rm Asy} \, \Gamma $ as
\begin{equation}
   {\rm Asy} \, \Gamma
   = \sum _{\pi '>\pi } \Gamma _{\pi '} + \Gamma _{\pi } +
       \mbox{other terms} .
\label{recursion}
\end{equation}
The ``other terms'' correspond to the three classes of surface
that are illustrated in Fig.\ \ref{fig:other.terms}:
\begin{itemize}

\item Those that are smaller than $\pi $.

\item Those that intersect $\pi $ in a subset (necessarily a
    manifold of lower dimension).

\item Those that do not intersect $\pi $ at all.

\end{itemize}
We assume as an inductive hypothesis that the sum of $\Gamma _{\pi '}$ over
$\pi ' > \pi $ gives a good approximation to the original $\Gamma $ except in
neighborhoods of the {\em smaller} surfaces for which $\Gamma _{\pi }$ has
not yet been constructed.  The integrals defining the $\Gamma _{\pi '}$'s
cover the whole of the space of integration variables, but they
are only required to give good approximations when one excludes
neighborhoods of smaller surfaces; more precisely we will require
them only to give good approximations when the test function in
Eq.\ (\ref{distrib}) has a zero of an appropriate strength on
these smaller surfaces.

We now construct $\Gamma _{\pi }$.
When combined with the $\Gamma _{\pi '}$ for larger
surfaces it must give a good approximation to $\Gamma $ on a
neighborhood of $\pi $.  It is sufficient for our purposes to
require only that we have a good approximation when the test
function has an appropriate zero on the smaller surfaces.  It is
not necessary to have constructed $\Gamma _{\pi '}$ for the smaller surfaces,
since they will give zero with such a test function.  This is
sufficient to prove the inductive hypothesis for the next use of
the recursion.

Since $\Gamma _{\pi }$ is localized on the surface $\pi $, it is necessary only
to consider a neighborhood of $\pi $.  This combined with our
remarks in the previous paragraph ensures that we do not need the
unconstructed ``other terms'' in Eq.\ (\ref{recursion}) in order
to construct $\Gamma _{\pi }$.

Therefore we define
\begin{equation}
   \Gamma _{\pi } = T_{\pi } \left(
              \Gamma  - \sum _{\pi '>\pi } R \Gamma _{\pi '}
            \right) ,
\label{asy.subs}
\end{equation}
where $T_{\pi }$ represents the Taylor expansion in powers of the small
variables on $\pi $.  The first term is the Taylor expansion of the
original graph, and the remaining terms can be thought of as
subtractions that prevent double counting of contributions to the
integral over a neighborhood of $\pi $.

The result is that a sum over $\Gamma _{\pi }$ and the terms for larger
regions,
\begin{equation}
   \Gamma _{\pi } +
   \sum _{\pi '> \pi } R \Gamma _{\pi '} ,
\end{equation}
correctly gives the contribution to the asymptotics of $\Gamma $ that
comes from a neighborhood of $\pi $ and of all larger regions, but with
neighborhoods of {\em smaller} regions being excluded.

Now, in general, $\Gamma _{\pi }$ gives a divergence when we integrate it
with a test function over a neighborhood of any of these smaller
regions.  So it is defined only when integrated with a test
function that is zero on these smaller regions.  We now extend it
to a distribution defined on all test functions by adding
infra-red counterterms to cancel the divergences.  (We will not
specify the details, but just observe that the construction is
exactly analogous to the construction of the well-known
distribution $(1/x)_{+}$.)  We call the result $R\Gamma _{\pi }$.
The counterterms are local in momentum space.
Since we have not yet considered how to
approximate $\Gamma $ in regions smaller than $\pi $, it is perfectly
satisfactory that we can {\em choose} a definition of
$\Gamma _{\pi }$ on the smaller
surfaces.  We only require that the result, $R\Gamma _{\pi }$, be finite, and
that the counterterms be localized on smaller surfaces than $\pi $,
so that we do not affect the good approximation we have already
obtained for $\pi $ and larger surfaces.

In the later stages of the recursion, we obtain the appropriate
approximations for these smaller regions. The subtraction terms,
as defined in Eq.\ (\ref{asy.subs}), ensure that changes in the
choice of counterterms localized on any particular surface $\pi $
are cancelled by corresponding changes in
the subtraction terms when we define $\Gamma _{\pi }$. Hence the overall
result for the asymptotic expansion of $\Gamma $ is independent of
these choices.

This completes the summary of the construction of the
${\rm Asy} \, \Gamma $.


\section{Completion of proof}
\label{sec:proof}

\subsection{Summary of previous results}

The results so far can be summarized in Eq.~(\ref{asy.basic}).
In the asymptotic large $Q$ limit, each graph is written
as a sum of contributions from a set of regions.  We have
identified the regions and computed the power of $Q$ associated
with each region.

Any particular region can be conveniently summarized by a diagram
of the form of Fig.\ \ref{fig:Reduced.Graph}.  It is specified by
a decomposition of a graph $\Gamma $ into two collinear subgraphs, $A$
and $B$, a soft subgraph, $S$, and a hard subgraph, $H$.  When we
sum Eq.~(\ref{asy.basic}) over all graphs $\Gamma $, we can represent
the result by independent summations over the possibilities for
the subgraphs $A$, $B$, $S$, and $H$:
\begin{equation}
   {\rm Asy} \, {\cal M} =
   \sum _{\Gamma } {\rm Asy} \, \Gamma  = A\times B\times S\times H .
\label{asy.summed}
\end{equation}
Here, $A$, for example, represents the sum over all possibilities
for a collinear-to-$A$ subgraph. Implicit in
Eq.~(\ref{asy.summed}) are appropriate Taylor expansions in small
variables, together with suitable subtractions to avoid double
counting, etc.  The symbol $\times $ represents integrations over the
momenta of loops that circulate between the different factors and
also a summation over the flavors of the parton lines joining the
different subgraphs.

Each subgraph comes with a specification of its external lines,
and the summation is restricted to compatible subgraphs.  For
example, in Fig.\ \ref{fig:Leading.Regions}(a) we require that
$H$
have as its external lines two collinear-to-$A$ gluons, a
collinear-to-$B$ quark, a collinear-to-$B$ antiquark, and the
virtual photon.  To be compatible with this, the subgraph $A$
must have as its external lines two collinear gluons, as well as
the hadrons $p$, $p'$ and the soft gluons.  Such restrictions can
be enforced by a suitable definition of the $\times $ operation in
Eq.~(\ref{asy.summed}).

\begin{figure}
    \begin{center}
        \begin{tabular}{c@{~~}c}
           \epsfxsize=0.2\hsize
           \epsfbox{lo-b.eps}
           \hspace*{0.2in}
           &
           \epsfxsize=0.2\hsize
           \epsfbox{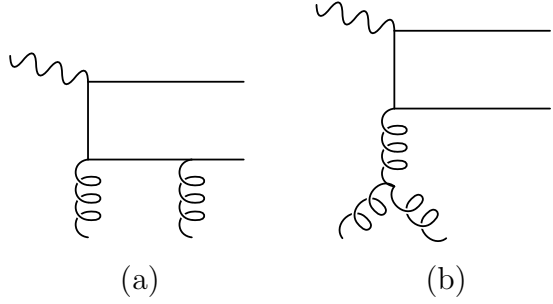}
           \hspace*{0.2in}
        \\
           (a) & (b)
        \end{tabular}
    \end{center}
\caption{(a) An allowed graph, and (b) a disallowed graph for a
    hard part whose external lines are two collinear-to-$A$
    gluons, a collinear-to-$B$ quark, a collinear-to-$B$
    antiquark, and a photon.}
\label{fig:H.graphs}
\end{figure}

As always for a hard subgraph, it is required that $H$ be
one-particle-irreducible (1PI) in the $A$ lines and the $B$
lines.  Thus, Fig.\ \ref{fig:H.graphs}(a) is allowed as a hard
subgraph.  However, Fig.\ \ref{fig:H.graphs}(b) is not allowed,
since it has an internal line (the vertical gluon) that is forced
to be collinear (by the two external gluons).

\subsection{Taylor expansion: collinear case}

\begin{figure}
    \begin{center}
        \leavevmode
        \epsfxsize=0.35\hsize
        \epsfbox{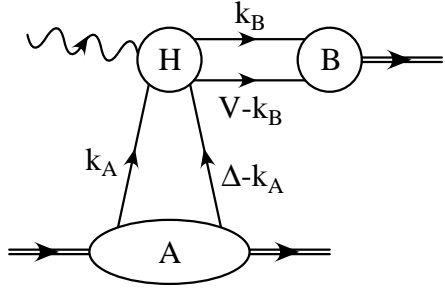}
    \end{center}
\caption{Simple region: collinear and hard subgraphs only, with
    two lines joining each collinear graph to the hard subgraph.}
\label{fig:ABH}
\end{figure}

We now Taylor expand the factors in Eq.~(\ref{asy.summed}) in
powers of small variables.  To understand the general principles
by which this operation gives the factorization theorem, with its
operator definitions of the collinear factors, let us first treat
the case that there is no soft factor and that exactly two lines
connect each collinear graph to the hard part---Fig.\
\ref{fig:ABH}.  We then have
\begin{eqnarray}
   A\times B\times H &=& \int  d^{4}k_{A} \, d^{4}k_{B}
                \, A(k_{A},\Delta -k_{A})
                \, B(k_{B}, V-k_{B})
                \, H(q, k_{A}, \Delta -k_{A}, k_{B}, V-k_{B})
\nonumber\\
         &\simeq& \int  d^{4}k_{A} \, d^{4}k_{B}
                \, A(k_{A},\Delta -k_{A})
                \, B(k_{B}, V-k_{B})
\nonumber\\
         &&     ~ H\!\left(q, (k_{A}^{+},0,0_{\perp }),
                     (\Delta ^{+}-k_{A}^{+},0,0_{\perp }),
                     (0,k_{B}^{-},0_{\perp }), (0,V^{-}-k_{B}^{-},0_{\perp })
                    \right) .
\end{eqnarray}
The notation is unfortunately cumbersome, but it makes precise
the operations we have applied to the hard part: We have replaced
the momenta collinear to $A$ by their $+$ components, and the
momenta collinear to $B$ by their $-$ components. This represents
the first term in the expansion of $H$ in powers of the other
components of these momenta.

Only the $k_{A}^{+}$ and $k_{B}^{-}$ integrals now couple the different
factors.  This gives
\begin{eqnarray}
   A\times B\times H  &\simeq& \int  dk_{A}^{+} \, dk_{B}^{-} \,\,
\nonumber\\
          &&     \int  dk_{A}^{-} \, d^{2}k_{A\perp } \,
                   A(k_{A},\Delta -k_{A}) \,\,
                 \int  dk_{B}^{+} \, d^{2}k_{B\perp } \, B(k_{B}, V-k_{B})
\nonumber\\
          &&    ~ H\!\left(q, (k_{A}^{+},0,0_{\perp }),
                     (\Delta ^{+}-k_{A}^{+},0,0_{\perp }),
                     (0,k_{B}^{-},0_{\perp }), (0,V^{-}-k_{B}^{-},0_{\perp })
                    \right) ,
\end{eqnarray}
which has the general structure of the factorization formula
Eq.~(\ref{factorization}).  To see this more explicitly, we
observe that:
\begin{itemize}

\item Scaling $k_{A}^{+}$ and $k_{B}^{-}$ by factors of $p^{+}$ and $V^{-}$,
   respectively, gives the integration variables $x_{1}$ and $z$
   in Eq.~(\ref{factorization}).

\item The factor $A$ is a matrix element of a time-ordered
   product of two fields.  Integrating over all $k_{A}^{-}$ and $k_{A\perp }$
   puts the difference of coordinates of the two fields on the
   light-like line $(0,y^{-},0_{\perp })$.  This is a matrix element of an
   operator like those in the definition of the parton density
   Eq.~(\ref{pdf.q.def}).

\item Similarly, the $B$ factor becomes like the meson wave
   function Eq.~(\ref{wf.def}).

\end{itemize}
At this point we have matrix elements of light-cone operators
that consist of two operators that are integrated along a
light-like line.

\subsection{Taylor expansion with soft factor; Glauber region}

To complete the proof, we now have to deal with the soft factor
in an analogous fashion and to show that the only operators we
need are the precise ones in the definitions
Eq.~(\ref{pdf.q.def})--(\ref{wf.def}). It is convenient to start
by considering $A\times S$ and $H\times B$ as units. Then we write
\begin{eqnarray}
  A\times B\times S\times H &=&
      \left( \prod _{i} d^{4}k_{i}\right)
      \, H\times B(q, V, k)
      \,\, A\times S(p, p', k)
\nonumber\\
   &=& \left( \prod _{i} d^{4}k_{i}\right)
       \, {\cal H}(q, V, k)
       \,\, {\cal A}(p, p', k) ,
\end{eqnarray}
where, the $k_{i}$'s are the loop momenta coupling the two factors.
Notice that ${\cal H} \equiv  H\times B$ is 1PI in the lines entering it
from ${\cal A} \equiv  A\times S$,
because any linear combination of momenta that are each collinear
to $A$ or soft is itself collinear to $A$ or soft.  On the other
hand, ${\cal A} = A\times S$ includes all graphs with the appropriate
number of external lines.

Clearly we may neglect the components $k_{i}^{-}$ of the soft momenta
within
${\cal H}=H\times B$, since by definition the momenta in both $B$ and
$H$ have $-$ components of order $Q$.  We may also neglect $k_{i\perp }$
within $H$.  But to derive the factorization theorem, we will
also need to neglect $k_{i\perp }$ within the $B$ subgraph.  In a general
situation this is not necessarily true, since the broadest
definitions of soft momenta and collinear-to-$B$ momenta only
insist that their $\perp $ components be small without specifying their
relative sizes. Hence one cannot always neglect a soft transverse
momentum with respect to a collinear transverse momentum.

\begin{figure}
    \begin{center}
        \leavevmode
        \epsfxsize=0.45\hsize
        \epsfbox{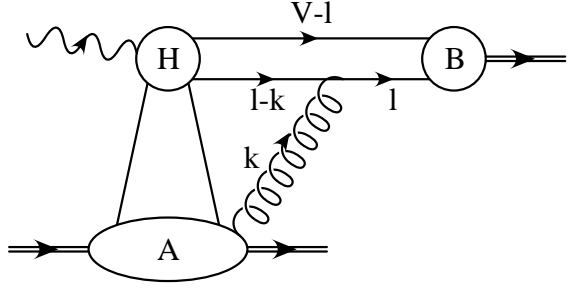}
    \end{center}
\caption{When $k$ is soft, this graph illustrates the need for
    contour deformation of $k^{+}$.}
\label{fig:Glauber}
\end{figure}

We use a version of the argument devised by Collins and Sterman
\cite{ColSt} for proving factorization for inclusive processes in
$e^{+}e^{-}$ annihilation.
The graph of Fig.\ \ref{fig:Glauber} illustrates the problem and
its solution.  We choose the gluon momentum $k$ to be soft, and the
quark momentum $l$ to be collinear to the meson.  The
momenta in the $A$, $B$, and $H$ subgraphs are, of course,
chosen to be collinear-to-$A$, collinear-to-$B$, and hard,
respectively.  Consider the integral over $k^{+}$, whose size is
much less than $Q$, since $k$ is soft.  For this reason, we
neglect $k^{+}$ in the subgraphs $A$ and $H$, and the only $k^{+}$
dependence is from the $S$ and $B$ subgraphs
\begin{eqnarray}
\lefteqn{
   \int _{{\rm soft~}k} dk^{+}
   \, \frac {1}{[(l-k)^{2}-m^{2}+i\epsilon ] \, (k^{2}+i\epsilon )}
}\hspace*{0.8in}
\nonumber\\
   &=&
   \int _{{\rm soft~}k} dk^{+} \,
   \frac {1}{[2(l^{+}-k^{+})(l^{-}-k^{-}) - (l_{\perp }-k_{\perp })^{2} - m^{2}
+ i\epsilon ] \, (2k^{+}k^{-} - k_{\perp }^{2} + i\epsilon )}
\nonumber\\
   &\simeq&
   \int  dk^{+} \,
   \frac {1}{[2(l^{+}-k^{+}) l^{-} - (l_{\perp }-k_{\perp })^{2} - m^{2} +
i\epsilon ] \, (2k^{+}k^{-} - k_{\perp }^{2} + i\epsilon )}
   ,
\label{glauber.eq}
\end{eqnarray}
where we have omitted inessential numerator factors.
In the second line of this equation, we have neglected $k^{-}$ with
respect to the large variable $l^{-}$.  Except for $l^{-}$, all the
momentum components used in this equation are small compared with
$Q$.

We distinguish two cases:
\begin{enumerate}

\item $k^{+}k^{-} \gtrsim k_{\perp }^{2}$.
    In this case, we can indeed neglect $k_{\perp }$
    in the first denominator.  Because $k$ is soft, while $l$ is
    collinear to $B$, the terms involving $k_{\perp }$ are small compared
    with the $k^{+}l^{-}$ term.

\item $k^{+}k^{-} \ll k_{\perp }^{2}$.  This is called the Glauber region in
the
    terminology of \onlinecite{BBL}.  In this region $k^{+}l^{-}$ may be
    comparable to $k_{\perp }^{2}$, so that we apparently cannot neglect
    $k_{\perp }$ in the collinear subgraph $B$.  However, in this region
    the only dependence on $k^{+}$ is in the collinear propagator,
    and so we may deform the $k^{+}$ contour into the complex plane
    until we recover the first case.

\end{enumerate}
So in fact we can neglect $k_{\perp }$ as well as $k^{-}$ in the collinear
propagator.

In general, we will have several soft momenta $k_{i}$ entering the
$B$ subgraph, and to use the above proof, we must ensure that
none of
the collinear propagators give obstructions to the contour
deformations for each $k_{i}^{+}$.  In other words, all the poles must
be on one side of the real axis for each $k_{i}^{+}$.  To prove this
\cite{ColSt},
we note that all the collinear-to-$B$ lines go forward from the
hard scattering, but not backward---compare the reduced graphs in
Fig.\ \ref{fig:Space.Time.2}.  Thus we can route all the $k_{i}^{+}$'s
back along collinear lines to the hard scattering, and thus all
the poles that collinear propagators give are in the
upper-half-plane, just as in Eq.~(\ref{glauber.eq}).

It should be observed that we cannot apply the same argument to
the $k^{-}$ dependence of the $A$ subgraph, since we have
collinear-to-$A$ lines both before and after the hard scattering.
This fact alone resulted in enormous complications in the proof
of factorization in the Drell-Yan process \cite{fact1,fact2}.

\subsection{Gauge-invariance}
\label{sec:ST}

Now that we have proved that the $+$ and $\perp $ components of soft
momenta may be neglected in both $B$ and $H$, we can
write\footnote{
    In this and the subsequent equations, the symbol `$\simeq$'
    means `equal up to power corrections'.
}
\begin{equation}
   A\times B\times S\times H \simeq
     \int  \prod _{i} dk_{i}^{+}
     \, {\cal H}(q, V, k^{+})
     \left( \int \prod _{i} dk_{i}^{-} d^{2}k_{i\perp }
        \,\, {\cal A}(p, p', k) .
     \right)
   \label{factored.1}
\end{equation}
This gets us much closer to the desired factorization.
It is exactly a kind of operator product expansion, since the
${\cal A}$ factor is a matrix element of a light-cone
operator, apart from the consequences of subtractions. In fact,
the subtractions needed to define ${\cal A}$ are associated with
regions with larger singular surfaces, and thus in fact to
ultra-violet divergences associated with the operator vertices.
That is, the subtractions are just an implementation of the
ultra-violet counterterms needed to define renormalized
operators.  We therefore write Eq.~(\ref{factored.1}) as
\begin{equation}
   A\times B\times S\times H \simeq
   \sum _{i} C_{i}(q,V,k^{+}) \, O_{i}(p,p',k^{+}) ,
\label{OPE}
\end{equation}
where the $O_{i}$ are the matrix elements of renormalized light-cone
operators, and we will call the $C_{i}$'s coefficient functions.  We
use $i$ to label the different possible operators.

But there are many possible operators, even when we restrict
ourselves to the leading power.  Each case of the graphs of Fig.\
\ref{fig:Leading.Regions} with a different set of external lines
for the $A\times S$ graph corresponds to a different operator.  But now
we can appeal to the new results by Collins \cite{NewST}.  These
show that we can restrict the sum to gauge invariant operators.
Such operators consist of gauge covariant operators (like $G_{\mu \nu }$,
$\psi $) joined by path-ordered exponentials (often called ``string
operators'').

We must now determine which of these operators is needed to give
a leading power. First, we construct a modified version of the
decomposition of gluons into scalar and transverse polarizations.
Consider one particular external gluon, of momentum $k$, that
attaches ${\cal A}$ to ${\cal H}$. We have a factor
\begin{equation}
   {\cal A}^{\mu }(k)  \, g_{\mu \nu } \, {\cal H}^{\nu }(k^{+}) ,
\end{equation}
where $g_{\mu \nu }$ is the numerator of the gluon propagator.
Recall, from Sec.\
\ref{sec:powers}, that the largest term in the sum over the
vector indices is the one with $\mu =+$ and $\nu =-$, i.e.,
${\cal A}^{+}{\cal H}^{-}$.
This happens because the collinear subgraphs are highly boosted
in the Breit frame and after the boosts the ${\cal A}^{+}{\cal H}^{-}$
term is the one with the largest components.  The arguments apply
both to the connection of collinear-to-$A$ lines to the hard
subgraph $H$ and of soft lines to the collinear-to-$B$ subgraph
$B$, i.e., to all the gluons connecting ${\cal A}$ to ${\cal H}$.

{}From the point of view of the ${\cal H}$ factor, the gluon $k$ is
an on-shell massless gluon with a polarization vector
proportional to ${\cal A}^{\mu }$, and a momentum in the $+$ direction:
$k'\equiv (k^{+},0,0_{\perp })$.
The big term in ${\cal A}\cdot {\cal H}$ therefore
corresponds to a polarization exactly proportional to the
momentum of the gluon. This we call a scalar gluon, and we
therefore make the following decomposition:\footnote{
   Notice that this definition has changed from the one we used
   earlier, Eq.~(\ref{gluon.pol.1}), in order to take account of
   the approximations we have made in the ${\cal H}$ subgraph.
}
\begin{equation}
   {\cal A}^{\mu } = p_{A}^{\mu }  \frac {{\cal A}\cdot p_{B}}{p_{A}\cdot
p_{B}}
        \, + \,
        \left({\cal A}^{\mu }
               - p_{A}^{\mu }  \frac {{\cal A}\cdot p_{B}}{p_{A}\cdot p_{B}}
        \right) .
\label{gluon.pol.2}
\end{equation}
To make a covariant formula, we used the previous definitions
that $p_{A}$ and $p_{B}$ are
vectors purely in the $+$ and $-$ directions.  The first term on
the right-hand-side of this equation we label as corresponding to
scalar polarization, and the second term as corresponding to
transverse polarization.  Since the scalar polarization is
exactly proportional to the approximated momentum $k'$ used in
${\cal H}$, it gives a factor $k'\cdot {\cal H}(k')$.
This is precisely the kind of
situation in which Ward identities simplify the sum over all
graphs.  The indirect methods of Ref.\ \onlinecite{NewST} give a very
efficient implementation of the relevant identities.

With the modified definitions, it is still true
that there is no penalty for attaching a scalar gluon to
${\cal H}$, but that there is a penalty for every transverse gluon
line and every quark line.  Now, the factors for the external
lines of ${\cal A}$ correspond to the Feynman rules for
light-cone operators.  So scalar gluons are associated with
factors of $A^{+}$ in an operator, where $A^{+}$ is the $+$ component
of the gluon field.
The gauge invariant gluon operator
with the lowest number of transverse gluons is of the form
\begin{equation}
   G^{i+}(0,y^{-},0_{\perp }) \, {\cal P} \, G^{j+} ,
\label{gluon.op}
\end{equation}
where ${\cal P}$ is a path-ordered exponential of the gluon
field.  The indices $i$ and $j$ label transverse components.
Notice that the operator associated with the scalar
gluons, $A^{+}$, is exactly the one that appears, exponentiated, in
${\cal P}$.

The operator, Eq.\ (\ref{gluon.op}), is a $2\times 2$ matrix on
transverse coordinates.  We now find the restrictions due to
angular momentum conservation that restrict those components of
this matrix that have a non-zero coupling to the hard scattering.
Consider the hard part times the meson
wave function as a scattering process for collinear gluons plus
the virtual photon to make the meson.  Angular momentum
conservation plus the fact that the photon is longitudinally
polarized implies that the angular momentum of the gluons around
the collision axis equals the helicity of the meson.

The matrix has components of helicities 0, $+2$ and $-2$.
So if the meson is a transversely polarized vector, then we have
a zero hard part, as indicated on the fourth line of Eq.\
(\ref{transverse.changes}).

If the meson is a longitudinally vector or a pseudo-scalar meson,
then either of the two matrices of zero helicity contribute:
\begin{equation}
    \left(
       \begin{array}{cc}
          1 & 0 \\
          0 & 1
       \end{array}
    \right)
    , \ \
    \left(
       \begin{array}{cc}
          0 & i \\
          -i & 0
       \end{array}
    \right)
    .
\label{gluon.matrix}
\end{equation}
Parity conservation implies that the first matrix is the only one
to which the hard scattering couples for the case of a
longitudinal vector meson.
For the factorization theorem for
longitudinal vector mesons, we therefore find that the gluon
density needed is the one defined in Eq.~(\ref{pdf.g.def}); it
defines the ${\cal A}$ factor, with the normalization factor being
a matter of convention.

For case of pseudo-scalar mesons, the second matrix in Eq.\
(\ref{gluon.matrix}) is the one that satisfies parity invariance.
However, charge-conjugation invariance, as indicated below Eq.\
(\ref{pseudo.scalar.changes}), implies that the hard-scattering
coefficient is zero.

Similar arguments give Eq.~(\ref{pdf.q.def}) as the definition
containing the smallest relevant gauge-invariant operator with
quarks, when we are treating production of longitudinally
polarized vector mesons, with Eq.\ (\ref{wf.def}) being the
definition of the meson wave function.
The $\gamma ^{+}$ factor in Eq.~(\ref{pdf.q.def}) picks out the
largest components of the quark and antiquark fields.

Next we apply the same arguments about angular momentum
conservation to the production of pseudo-scalar mesons and of
transversely polarized vector mesons. We find that the
changes needed in the definitions of the parton densities and the
wave functions are those indicated in Eqs.\
(\ref{pseudo.scalar.changes}) and (\ref{transverse.changes}).

In our expansion of the form of Eq.~(\ref{OPE}), the operators
can be expanded in powers of the fields in the path-ordered
exponentials.  Thus we may regard Eq.~(\ref{OPE}) in two
equivalent ways.  One way is to restrict the operators to exactly
the gauge invariant operators.  The number of terms is then
$2N_{f}+1$: one operator
for each flavor of quark and antiquark and one for
the gluon.  Another way to look at the formula is to sum over
terms for each of the operators obtained in the expansion of the
gauge-invariant operators.  This gives an infinity of terms, but
in $2N_{f}+1$ groups, with identical Wilson coefficients within each
group.

The second point of view is useful because it shows that, to
obtain the coefficient $C_{i}$ for each gauge-invariant operator,
it is sufficient to examine graphs for the $H\times B$ factor that have
the minimum number of external lines, viz., two transverse gluons
or two quarks (in addition to the photon and the meson).

\subsection{Endpoint contributions}
\label{sec:endpoint}

We have now shown that the leading power for our amplitude is
given by a convolution of operator matrix elements for the
proton, times coefficients that are obtained from hard subgraphs
times collinear subgraphs associated with the meson.  The
coefficients are obtained from graphs for ${\cal H} = H\times B$ with
exactly the two external lines that correspond to the two parton
fields in Eqs.~(\ref{pdf.q.def}) and (\ref{pdf.g.def}) that we
have when the path-ordered exponentials are omitted in the
operators.

Only if both external lines connect to the hard part can we
proceed to the next step of factoring $H\times B$ into the hard factor
in the factorization formula times the wave-function factor
Eq.~(\ref{wf.def}).  Unfortunately, the external lines of $H\times B$
can connect either to the hard subgraph or to the collinear
subgraph, a situation summarized in the following equation:
\begin{equation}
    \raisebox{-0.25in}{\epsfbox{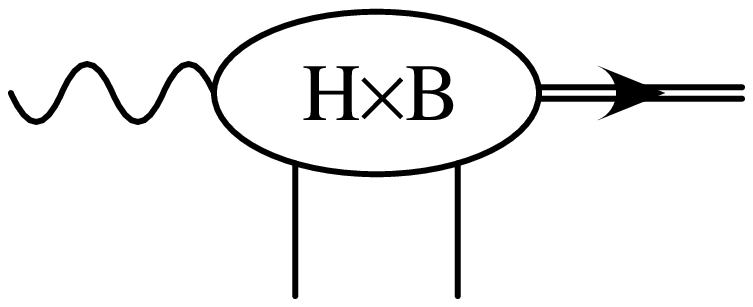}}
    ~=
    \raisebox{-0.25in}{\epsfbox{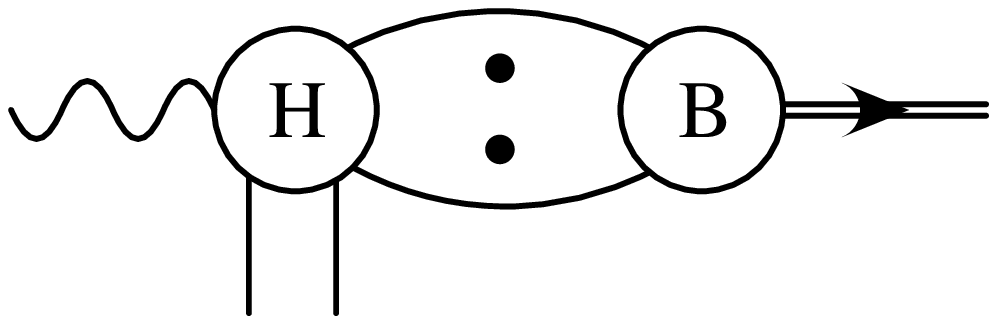}}
    ~+
    \raisebox{-0.25in}{\epsfbox{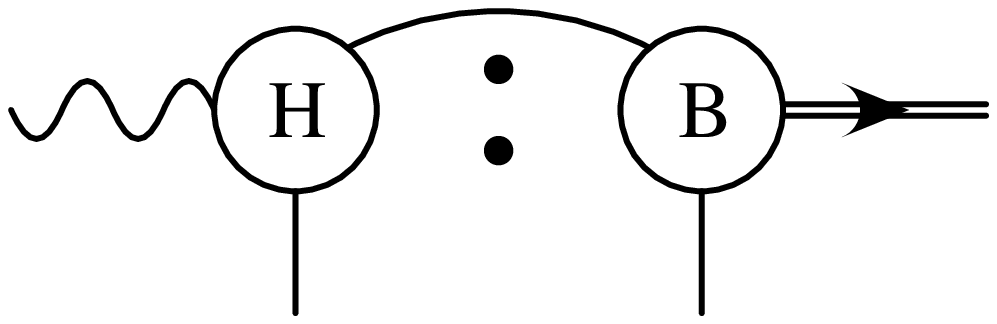}}
    .
\label{HB.eq}
\end{equation}
The first term corresponds to the region of Fig.\
\ref{fig:Leading.Regions}(b), and the second to Fig.\
\ref{fig:Leading.Regions}(f), in the case that there are no extra
gluons attaching $S$ to $B$.  The dots between the $H$ and $B$
subgraphs indicate an arbitrary number of lines being exchanged.

A similar equation applies with external gluons, and corresponds
to the regions of Fig.\ \ref{fig:Leading.Regions}(a), (d), and
(e). Note that the gluon attaching to $B$ now has to be
transverse, so that we have lost one power of $Q$ for graph (e),
which has only three partons connecting to the hard subgraph.
This brings the power for all cases down to $1/Q$, and hence
there are no further power law cancellations that we will need to
take into account.

In Eq.~(\ref{HB.eq}), we call the term where one of the lines
attaches to $B$ an ``end-point'' contribution, for the following
reason.  In the factorization equation (\ref{factorization}), the
longitudinal momentum fractions of the two lines relative to the
incoming proton are $x_{1}$ and $x-x_{1}$.  When one of the lines
attaches to $B$, that means that the line is soft, that is, that we
are examining the contribution of a small neighborhood of either
$x_{1}=0$ or $x-x_{1}=0$.  We can equally well think of the
contribution as being obtained from a region of the form of Fig.\
\ref{fig:Leading.Regions}(a) or (b), when one of the quarks
joining the meson to the hard part becomes soft.  That is, the
term can also be thought of as related to one of the endpoints
$z=0$ or $z=1$ of the $z$ integral.

{\em Suppression of endpoint contribution for longitudinal
photon:}  According to our power counting formula, the end-point
contribution is leading, being proportional to $1/Q$.  There is
in fact an additional suppression.  Consider Fig.\
\ref{fig:Leading.Regions}(e).  The hard part is proportional to
\begin{equation}
   \epsilon _{g}^{i} \epsilon _{\gamma }^{\mu } h_{i\mu } ,
\end{equation}
where $\epsilon _{g}$ and $\epsilon _{\gamma }$ are the polarizations of
the gluon and
photon, and the gluon index $i$ is purely transverse.  The tensor
$h_{i\mu }$ is obtained from the diagrams with a trace with $\gamma ^{+}$.
Thus $h_{i\mu }$ is a tensor constructed from
vectors in the $+$ and $-$ direction and from the metric tensor.
It is therefore zero when $g=+$ or $\mu =-$, and therefore we get
a zero when the photon is longitudinally polarized.  The
proofs we have made previously are appropriate for the leading
power of $Q$, so the zero corresponds to a suppression by another
power of $Q$.  Recall that we have already proved that at least
one of the gluons joining $S$ to $B$ must be transverse, and that
results in a suppression compared with the $Q^{0}$ given by
the power-counting formula.

We therefore conclude that the endpoint contribution from
Fig.\ \ref{fig:Leading.Regions}(e) is of order $1/Q^{2}$.
Since we have already proved---around Eq.~(\ref{QuarkFF})---the
corresponding result for Fig.\ \ref{fig:Leading.Regions}(f),
where the hard scattering has two quark lines, we now know that
all endpoint contributions are suppressed, and we saw above that
this is sufficient to obtain the factorization theorem.

But clearly, the situation is different when the photon is
transversely polarized.  We will discuss this further in Sec.\
\ref{sec:transverse}.

\subsection{End of proof}

We have now proved that the endpoint contributions are
power-suppressed, in the case that the photon has longitudinal
polarization.  So the only term that survives in
Eq.~(\ref{HB.eq}) is the one where both partons from the proton
attach to the hard scattering.

We can now apply the operator expansion argument to the $H\times B$
factor, to obtain the product of a coefficient times a suitable
vacuum-to-meson matrix element.  The matrix element is the one
given in Eq.~(\ref{wf.def}), with no purely gluonic operator
being allowed, because of our choice for the quantum numbers of
the meson.  This result immediately gives the factorization
theorem, Eq.~(\ref{factorization}), provided only that we adjust
the normalization of the hard-scattering factor appropriately.


\section{Evolution equations}
\label{sec:evolution}

The definitions of the off-diagonal parton densities,
Eq.~(\ref{pdf.q.def}) and (\ref{pdf.g.def}), are just the same as
those of the ordinary diagonal parton densities.  In both cases,
there are ultra-violet divergences and corresponding anomalous
dimensions. The divergences are properties of the operators
themselves.  Since the same operators appear in the light-cone
wave function, Eq.~(\ref{wf.def}), this permits us to give a
unified treatment for both the parton densities and the wave
functions.

The resulting renormalization-group equations give the DGLAP
evolution that is essential to phenomenology.  The two
non-perturbative factors in the factorization theorem depend on a
renormalization/factorization scale $\mu $.  We need to choose it of
order $Q$ in order to make effective perturbative calculations of
the hard scattering factor.  Therefore we need the evolution
equations with respect to $\mu $, in order to compute predictions in
terms of the non-perturbative factors evaluated at a fixed scale.

Only minor generalizations in previously existing treatments for
the diagonal densities are needed \cite{pdfs}.  Balitsky and
Braun \cite{Balitsky.Braun} have given a more general treatment,
and recently Ji and Radyushkin \cite{Radyushkin2,Ji,Radyushkin1}
treated exactly the operators we are considering.

\begin{figure}
    \begin{center}
        \leavevmode
        \epsfbox{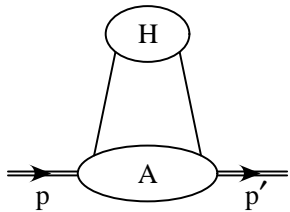}
    \end{center}
\caption{Regions for UV divergences of parton densities.  The
         momenta in the upper blob have large virtualities, and
         the momenta in the lower blob are collinear to the
         hadron. Removing one external hadron gives the regions
         for UV divergences of light-cone wave functions.}
\label{fig:UV.div}
\end{figure}

The essential point is that the ultra-violet divergences arise
when momenta get infinite in a subgraph attached to the operator
vertex.  The relevant regions of loop momentum can be labeled by
diagrams of the form of Fig.\ \ref{fig:UV.div}, which is to be
interpreted in a similar fashion to those for the leading regions
for the scattering amplitude itself.

After use of gauge invariance to make a kind of operator product
expansion, the divergences will be of the form of the parton
densities themselves convoluted with ultra-violet renormalization
factors.  Hence the right-hand side of the evolution equation is
of the form of a kernel convoluted with the parton densities.
The derivation and the result is just the same as for the
diagonal densities, except that one must
take account of the longitudinal momentum flow in the $t$-channel.
For the distributions, we have
\begin{equation}
   \mu \frac {d}{d\mu } f_{i/p}(x_{1}, x_{2}, t, \mu ) =
   \sum _{j} \int  d\xi
   P_{ij}(x_{1}, x_{2}, \xi ; \alpha _{s}(\mu ))
   f_{j/p}(\xi , x_{2}-x_{1}+\xi , t, \mu ) .
\label{AP.distrib}
\end{equation}
When $t=0$ and $x_{1}=x_{2}=x$, the
equation reduces to the standard Altarelli-Parisi equation, with
a kernel $\xi P_{ij}(x, x, \xi ; \alpha _{s}(\mu ))$.
Since the ultra-violet
divergences are independent of the transverse and
$-$ components of momenta, the kernel $P_{ij}$ is independent of
$t$.

This implies that when the individual momentum fractions $x_{1}$ and
$x_{2}$ are much larger than $x\equiv x_{1}-x_{2}$, the distributions approach
the diagonal ones, and the limit $x_{1}=x_{2}$ can be taken in the
kernel.

The same operator occurs in the meson's light-cone wave function,
so that its evolution equation contains the same kernel
\begin{equation}
   \mu \frac {d}{d\mu } \phi ^{V}_{i}(z, \mu ) =
   \sum _{j} \int  d\zeta
   P_{ij}(z, \zeta ; \alpha _{s}(\mu ))
   \phi ^{V}_{j}(\zeta , \mu ) .
\label{AP.wf}
\end{equation}

Corresponding Altarelli-Parisi equations apply to the other
parton densities and wave functions needed for treating the
production of pseudo-scalar mesons and transversely polarized
vector mesons.

\section{Rules for Hard scattering function}
\label{sec:hard}

The hard scattering function $H_{ij}$ in Eq.~(\ref{factorization})
is obtained from graphs with the appropriate external parton
lines for the $H$ subgraph Fig.\ \ref{fig:Leading.Regions}(a)
and (b).   The graphs are 1PI in the two lines from the proton
and in the two lines from the meson.  Lowest order graphs are
given in Fig.\ \ref{fig:LO.Graphs}.  Subtractions are made to
cancel the collinear divergences.  Minimal subtraction can be
used for the subtractions just as in inclusive hard scattering,
and in the same fashion. Normal Feynman rules are applied to the
interior of the graphs, so it remains to construct the
normalization factors and the external line factors.

Consider first graphs in which the proton factor is connected by
quark lines to the hard scattering. The leading power is obtained
from a factor of the form
\begin{equation}
   {\rm tr} \, \left(
      \mbox{$\gamma ^{+}$ part of $H$}
      \times  \mbox{$\gamma ^{-}$ part of $A$}
   \right) ,
\end{equation}
which we can write as
\begin{equation}
   \frac {1}{2}{\rm tr} \, (\gamma ^{-}H)
   \,
   \frac {1}{2}{\rm tr} \, (\gamma ^{+}A) .
\end{equation}
The $1/2$ and the $\gamma ^{+}$ in the second factor appear directly in
the definition of the parton density, Eq.~(\ref{pdf.q.def}), with
the $1/2$ multiplying a $1/2\pi $ associated with the Fourier
transform.

As to the integral over the loop momentum $k$ connecting $A$ and
$H$, the $1/(2\pi )^{4}$ factor is completely inside the parton
density, as are the integrals over the transverse and $-$
components of the momentum.  We rewrite the integral over $k^{+}$
as
\begin{equation}
   \int  dk^{+} \, \dots
   = \int  dx_{1} \, p^{+} \dots .
\end{equation}
In addition there is a trace over color indices between $A$ and
$H$.  Since $A$ is a unit matrix in color space---the protons are
color singlet---and since the parton density is defined to
include a sum over colors, we need to trace $H$ over color and
divide by the number of quark colors, $N_{c}=3$.

Hence the external line factor associated with quarks entering
the hard scattering from the proton blob is
\begin{equation}
   \mbox{$A$-quark factor}
   = \frac {1}{2N_{c}} p^{+} \,\, {\rm tr} \, \gamma ^{-} \dots ,
\label{A.q.factor}
\end{equation}
with the trace being over both Dirac and color indices, and where
``$\dots$'' represent the rest of the hard subgraph, with
ordinary Feynman rules.
The factor $1/2N_{c}$ is in effect an average over spin and color,
just as we would have in an inclusive process.

With one exception,
exactly similar considerations apply to the connection of the
hard scattering to the meson factor, apart from a need to
exchange the $+$ and $-$ coordinates.   The exception
is that the definition Eq.~(\ref{wf.def}) of the wave function
contains an extra factor $1/\sqrt {2N_{c}}$. Hence the external line
factor associated with quarks entering the hard scattering from
the meson blob is
\begin{equation}
   \mbox{$V$-quark factor}
   = \frac {1}{\sqrt {2N_{c}}} V^{-} \,\,
   {\rm tr} \, \gamma ^{+} \dots .
\label{B.factor}
\end{equation}

Finally there is the case of gluons attaching the proton blob to
the hard subgraph.  Here we get similarly
\begin{equation}
   \mbox{$A$-gluon factor}
   = \frac {1}{2(N_{c}^{2}-1)} \delta _{ij} ,
\label{A.g.factor}
\end{equation}
where we have an average over the two transverse polarizations and
the $N_{c}^{2}-1$ colors of a gluon.


\section{Transversely polarized photons}
\label{sec:transverse}

Our proof of the factorization theorem Eq.~(\ref{factorization})
is valid when the photon is longitudinally polarized, since we
were able to show that the contribution of endpoint regions was
power suppressed. Order-by-order in perturbation, an amplitude of
order $1/Q$ times logarithms was obtained, but with an
enhancement due to scaling violations when we apply DGLAP
evolution.

In this section we will show that the amplitude for transversely
polarized photons is suppressed by one power of $Q$.  First we
will show this for the non-endpoint contribution, as a
consequence of Lorentz invariance.  Then we will treat the
endpoint terms.

\subsection{Power-counting: non-endpoint case}
\label{sec:transverse.non.endpoint}

Consider first the non-endpoint contributions, where the hard
scattering has four external lines, Fig.\
\ref{fig:Leading.Regions}(a) and (b).  For region (a), the hard
part has a polarization-dependence of the form
\begin{equation}
   \epsilon _{1}^{i} \epsilon _{2}^{j}
   \epsilon _{\gamma ^{*}}^{\mu } h_{ij\mu } ,
\label{3index}
\end{equation}
where $\epsilon _{1}$ and $\epsilon _{2}$ are the transverse gluon
polarizations, and
$h_{ij\mu }$ is a tensor constructed out of longitudinal vectors and
out of Lorentz invariants.  The tensor is therefore invariant
under rotations in the transverse plane, and hence it can only be
nonzero if $\mu =+$ or $\mu =-$.  So we get a non-zero result for
the leading power only for a longitudinally polarized photon.

For the quark graph, \ref{fig:Leading.Regions}(b), the result is
even simpler, since after the trace over Dirac matrices, the hard
part just gives a vector
\begin{equation}
   \epsilon _{\gamma ^{*}}^{\mu } h_{\mu } .
\label{2index}
\end{equation}
The same argument that we applied to Eq.~(\ref{3index}) gives
exactly the same result.

Therefore in the case of the non-endpoint contribution
the leading $1/Q$ power is only obtained
when the photon is longitudinally
polarized.  There must be at least a $1/Q$ suppression for
transversely polarized photons, which gives a final power $1/Q^{2}$.
Now in Sec.\ \ref{sec:endpoint}, we showed that the endpoint
terms obey exactly the opposite rule: longitudinal photons are
suppressed, and transverse photons give the $1/Q$ contributions.

\subsection{How soft is soft?}
\label{soft.soft}

However, to get the $1/Q$ contribution with a transverse photon,
we depend on the soft momenta being treated as having a magnitude
of $m^{2}/Q$.  This is evidently very small: the corresponding
virtuality is of order $m^{4}/Q^{2}$. Clearly, we must expect
non-perturbative confinement effects to restrict all significant
virtualities to being $m^{2}$ or larger. We now show that this
results in a power suppression.

\begin{figure}
    \begin{center}
        \leavevmode
        \epsfbox{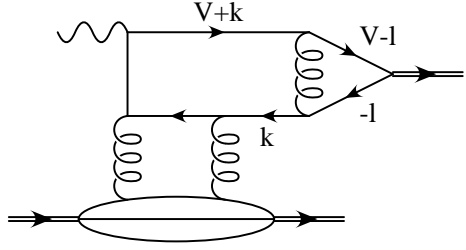}
    \end{center}
\caption{Graph to illustrate endpoint contribution with
    transversely polarized photon.}
\label{fig:endpoint.example}
\end{figure}

To see what is happening, let us examine a particular graph,
Fig.\ \ref{fig:endpoint.example}.  There are many ways in which
regions of the form of Fig.\ \ref{fig:Reduced.Graph} can be
constructed.  For our purposes it will be sufficient to restrict
our attention to cases where the momentum $l$ that goes through
the meson vertex is always collinear to $B$ and not close to
either of its endpoints. Similarly, we choose all momenta in
subgraph $A$ to be collinear to $A$.

The region of interest for this purpose is where the loop momentum
$k$ becomes soft.  So let us suppose that all components of $k$
are of order $\lambda Q$, where $\lambda $ is a parameter that we will vary
between $m^{2}/Q^{2}$ and unity.  The upper limit of this range is
where $k$ becomes a hard momentum, and the lower limit is where
$k^{\pm }$ become comparable to the small components of collinear
momenta.  Thus when $\lambda $ is outside of these limits we get
a power-law suppression.

All the propagators and loop integrals give factors of order
unity except for those in the loop $k$.  So we just need to focus
our attention on the following factor:
\begin{eqnarray}
   \int  d^{4}k ~ {\rm tr} \,
   &&
   \left[
      \slash \epsilon _{\gamma ^{*}}
      \frac {\slash V + \slash k}{(V+k)^{2}}
      \frac {1}{(k+l)^{2}} \gamma ^{\mu } \slash B \gamma _{\mu }
      \frac {\slash k}{k^{2}}
      \slash \epsilon _{1}
      \frac {\slash A_{1} + \slash k}{(A_{1}+k)^{2}}
      \slash \epsilon _{2}
      \frac {\slash A_{2} + \slash k}{(A_{2}+k)^{2}}
   \right] .
\label{example}
\end{eqnarray}
Here $A_{1}$ and $A_{2}$ represent two collinear momenta associated
with the two lower gluons, while $\epsilon _{1}$, $\epsilon _{2}$
and $\epsilon _{\gamma ^{*}}$ are the polarizations of
these two gluons and of the
photon.  We use $B^{\mu }$ to denote a collinear vector associated with the
right-hand-loop through the meson.
In $(+,-,\perp )$ coordinates, the magnitudes of the $A$ and $B$
momenta are
\begin{eqnarray}
    A_{1}^{\mu }, A_{2}^{\mu } &\sim& \left(
                       Q, \frac {m^{2}}{Q}, m
                    \right) ,
\\
    B^{\mu } &\sim& \left(
                 \frac {m^{2}}{Q}, Q, m
               \right) .
\end{eqnarray}

{\em Hard region for $k$:}
When $\lambda  \sim 1$, so that all components of $k$ are of order $Q$,
we get an overall power $1/Q$ made up as follows:
\begin{itemize}

\item $Q^{4}$ for the integration $d^{4}k$.

\item $1/Q^{10}$ for five hard denominators.

\item $Q^{5}$ for the five numerators, each of which has at least
    one term of order $Q$.

\end{itemize}
As always, we are working in the Breit frame.  The power $1/Q$ is
exactly what we obtained from general arguments; the hard
subgraph consists of the $k$ loop and has external line four
partons
and the photon.  Given the cancellations proved in Sec.\
\ref{sec:ST}, we know that we can take the gluon polarizations
to be transverse; this fact was used in obtaining the power of
$Q$ for the numerator.  Furthermore, the argument in Sec.\
\ref{sec:transverse.non.endpoint}
shows that after the integration over the
azimuth of $k_{\perp }$, the power $1/Q$ is only obtained when the photon
has longitudinal polarization; one power of $Q$ is lost for
transverse polarization.

{\em Soft region for $k$:}
Next we consider smaller values of $\lambda $.  There are two ranges to
consider:
$1 > \lambda  > m/Q$ and $m/Q > \lambda  > m^{2}/Q^{2}$.
The breakpoint $\lambda =m/Q$ between the two ranges occurs where the
components of $k$ are comparable to masses and typical collinear
transverse momenta.

In the higher range $1 > \lambda  > m/Q$ we obtain a power $\lambda /Q$, as
follows:
\begin{itemize}

\item $\lambda ^{4}Q^{4}$ for the integration $d^{4}k$.

\item $1/(\lambda ^{4}Q^{8})$ for the four denominators of the form
    $\left(\mbox{Collinear momentum} + k \right)^{2}$.

\item $1/(\lambda ^{2}Q^{2})$ from the $k^{2}$ denominator.

\item $\lambda ^{3}Q^{5}$ for the five numerators.

\end{itemize}
The numerator is the product of a $\slash k$ factor, which is of
order $\lambda Q$, and of four factors each with a largest component of
order $Q$.  But the large components of collinear momenta are in
a light-like direction.
Since $(\gamma ^{+})^{2} = (\gamma ^{-})^{2} = 0$, we cannot be
restricted to just the biggest terms in the momenta, and
examination of the surviving terms shows that the result for the
numerator is in fact $\lambda ^{3}Q^{5}$.

Note that the two lines of momenta $A_{1}+k$ and $A_{2}+k$ are
off-shell by much more than of order $m^{2}$.  Thus they are hard
relative to the collinear gluons and the argument that the gluons
are transverse still holds.  We do not have to be concerned about
a scalar gluon polarization.

The overall result, $\lambda /Q$, is correct if the photon has
longitudinal polarization.  If the photon is transverse, the
power is in fact $1/Q^{2}$.  This can be seen on an examination of
the trace algebra by noting that the number of gamma matrices in
the transverse direction must be even, and that after an
azimuthal average over $k_{\perp }$, the number of factors of $k_{\perp }$ must
be even.  The transverse $\slash \epsilon _{\gamma ^{*}}$
must be balanced by using
the transverse part of $\slash B$, which is of order $\lambda ^{0} Q^{0}$.
This results in replacing one factor of $\lambda Q$ by unity.

Hence the amplitude for a transverse photon is smaller than
the amplitude for a longitudinal photon until the lower end of
the region we are considering, at $\lambda  \sim m/Q$.  In any event we
always have a power suppression compared to the dominant part of
the amplitude with a longitudinal photon.

{\em Super-soft region for $k$:}
The situation changes once $\lambda $ goes below $m/Q$.  In the real
world, we must
suppose that this region, which we will call the `super-soft
region,' is suppressed due to confinement effects.  We could
model such effects within perturbation theory by giving the
partons non-zero masses.  But as an exercise, it is instructive
to obtain the size of the contribution when the partons
have zero masses.

First we observe over the whole of this region,
$m/Q > \lambda  > m^{2}/Q^{2}$, the power counting for the range of
integration and the denominators remains true, to give
$1/(\lambda ^{2}Q^{6})$; all the changes are in the numerator factor.  The
numerator is a sum of terms each of which is the product of five
individual momentum components. The biggest terms have two
factors of $Q$ from the large components of the collinear
momenta, and one factor of $\lambda Q$ from $\slash k$. In the remaining
two factors, the largest components are of order $m$ instead of
$\lambda Q$.  (Here is one motivation for separating the two parts of
the soft region at $\lambda  \sim m/Q$.)
Hence the numerator must be treated as begin
of order $\lambda Q^{3}m^{2}$ to give a total power $1/(\lambda Q^{3})$.
Note that this term only exists for transverse photons, as we
proved at the end of Sec.\ \ref{sec:catalog}; it is power
suppressed for longitudinal photons.

At the lower end of the region, $\lambda  \sim m^{2}/Q^{2}$, we obtain a
leading power contribution.

\subsection{Summary of results for transversely polarized photon}

We have shown that for a transversely polarized photon, there is a
suppression of $1/Q$ in the amplitude relative the the case of a
longitudinal photon. Now we discuss the significance of this, and in
particular the apparent lack of a simple factorization theorem,
and of a simple parton model interpretation of the results.

For the non-endpoint contribution, the suppression results from the
properties of the Dirac traces. For example, in Eq.~(\ref{example}) we
cannot replace all the factors in the trace by their largest components
without obtaining the trace over an odd number of transverse Dirac
matrices. The $1/Q^{2}$ contribution is obtained by replacing one of the
matrices by an order $m$ term instead of an order $Q$ term. This may
involve either circulating transverse momentum in the hard subgraph, or
a replacement of $\slash B$ by a transverse part. In either case,
the operators needed to define the collinear factors are no
longer the ones in the definitions of the parton densities and
wave functions, Eqs.~(\ref{pdf.q.def})--(\ref{wf.def}), since we
need to project out different components of Dirac matrices and/or
define an operator sensitive to parton transverse momentum.
Hence the factorization theorem Eq.~(\ref{factorization}) does
not hold, even when we restrict attention to the the non-endpoint
contribution; with a transverse photon, we must not only change
the hard scattering factor but we must also put in more general
objects for the non-perturbative factors.

For the endpoint contribution, we have to allow for a
non-perturbative soft factor.  Just as in the case of the
transverse-momentum distribution for the Drell-Yan and other
processes \cite{DY}, we should be able to do this by defining a
suitable phenomenological function to be convoluted with the
other factors in the amplitude.  It would be an interesting
result to derive a general result beyond leading logarithm
approximation.

In any event the results for transverse photons appear to be more
complicated and difficult than for longitudinal photons.  It is
not possible to use a {\em naive} generalization of the
factorization theorem we have derived with a longitudinal photon.

Previous work \cite{BFGMS,FKS,Ryskin} on this process has used
the proton rest frame rather than the photon rest frame.
Although that is a useful frame for deriving leading logarithm
results, and for gaining intuition about how the process works,
it is not so useful in constructing a complete factorization
theorem.  However, it is worth noting the corresponding results.
We let $z$ be the momentum fraction carried by the quark joining
the meson to the hard scattering.  Then $z$ is very similar to
the parameter $\lambda $ we used in investigating the endpoint
contribution, in Sec.\ \ref{soft.soft}.  The endpoint
contribution arises when $z$ is close to $0$ or $1$.  If $z$ is
of order $m/Q$ a $1/Q^{2}$ contribution for the amplitude was
obtained with a transverse photon \cite{BFGMS}, and if $z$ gets
unphysically small, of order $m^{2}/Q^{2}$, we get a $1/Q$
contribution.  There are additional Sudakov suppressions when
$Q^{2}$ is large enough.


\section{Predictions for relations between cross sections for
          different mesons}
\label{flavor.dependence}

As in the case of inclusive processes, the factorization theorem
leads to predictions for the flavor dependence, in this case for
relations between the cross section for productions of mesons of
different flavors.

\subsection{Small $x$}

At small $x$, the parton densities are dominated by exchange of
vacuum quantum numbers, since this is just a normal Regge limit.
Thus to a good approximation the factor of the hard scattering
times the parton density will be proportional to the square of
the charge of the quark connecting the hard subgraph to the
meson.  If we now make the approximation that the wave functions
for the different mesons,
$\rho ^{0}$, $\omega $, $\phi $ and $J/\psi $, are the same
apart from the obvious flavor dependence, we get the prediction
\cite{FKS} that their production cross sections are in the ratios
\begin{equation}
   \rho ^{0} : \omega  : \phi  : J/\psi
   = 9 : 1 : 2 : 8.
\label{ratios}
\end{equation}
We should expect this approximation to be reasonable for the
three
light mesons, but not so good for the $J/\psi $.
Since the $J/\psi $ is smaller than the light mesons, we should
expect its production cross section to be even larger than
predicted by this formula.
[The particular prediction Eq.\ (\ref{ratios}) for
the $J/\psi $ also depends on $Q^{2}$ being large enough that the
charmed quark mass can be neglected in the hard scattering.]

What the results of this paper give is that the prediction
Eq.~(\ref{ratios}) is immune to higher order QCD corrections.
That is, its accuracy only depends on the use of small $x$ and on
the similarity of the meson wave functions.

\subsection{Large $x$}

At large $x$, the dominant parton flavors in the proton are the
valence quarks.  Although we do not know the non-diagonal parton
densities, it is highly likely that they will be qualitatively
similar to the diagonal densities.
In particular, the biggest will be those for the $u$ and $d$
quarks, and the $u$ density in a proton will be rather bigger
than density of $d$ quarks.
So the production of the $\phi $ and $J/\psi $ mesons
will be suppressed compared with the values at small $x$.  Also
the fact that there are fewer down than up quarks will reduce the
suppression of $\omega $ production.  We see this as follows.  Let the
meson wave function have flavor dependence of the form
$a \bar u u + b \bar d d$, and let $R$ be the ratio of up to down
quarks in the proton.  Recall that for the $\rho ^{0}$, $a=-b$, whereas
for the singlet $\omega $, we have $a=+b$.

Let us work to lowest order in the hard scattering and ignore the
small gluon contribution.  Then the cross section is proportional to
$(2Ra-b)^{2}$, since the lowest-order hard scattering amplitude
depends on quark flavor only through a factor of the quark
charge.  Notice that there is interference in the amplitude
between the terms with different flavors of quark, and we have
destructive interference for the singlet $\omega $.

The ratio of cross sections is
\begin{equation}
   \rho ^{0} : \omega   =  (2R+1)^{2} : (2R-1)^{2} .
\end{equation}
When $R=1$ this
is the $9:1$ ratio in
Eq.~(\ref{ratios}).  When $R$ increases above $1$, the $\rho ^{0}/\omega $
ratio gets a lot closer to unity.  For example, when $R=2$
which is natural for $x \sim 0.2$
we get
a ratio of $2.8 : 1$.

We also observe that the $\phi /\rho $ ratio should decrease with
increasing $x$, compared with Eq.\ (\ref{ratios}), which assumed
vacuum quantum number exchange.  The decrease results from the
lack of strange quarks in the proton. This may be relevant for
the significantly smaller $\phi /\rho $ ratio that is observed at
the New Muon Collaboration (NMC)
compared to the DESY collider HERA
at similar $Q^{2}$.

\subsection{Production of transversely polarized vector mesons}

The production of transversely polarized vector mesons involves
the quark transversity density $\delta q$ (or $h_{1}$).  Normally one
would imagine that at small $x$, such parton densities are a
power of $x$ smaller than the regular, unpolarized parton
densities, and in particular than the gluon density.  This is
because the transversity density involves a helicity flip.  It
is usually expected that this requires exchange of non-vacuum
quantum numbers, whereas small $x$ physics is dominated by
something like Pomeron exchange.

Thus the ratio of transversely polarized vector mesons to
longitudinally polarized vector mesons should be small at small
$x$, and go to zero at $x=0$.  Thus we are unable to explain the
reported ratio from ZEUS: $\sigma _{L}/\sigma _{T}=1.5{+2.5 \atop
-0.6}$\cite{ZEUS}, since the ZEUS data are at small $x$, around
$10^{-2}$.  It is possible that the $Q^{2}$ of the data is small
enough that there is significant production by transversely
polarized {\em photons}.  The selection rules in this case are
different, and one need not have the same suppression of
transverse polarization for the meson.

On the other hand, there is no reason for the same suppression at
large $x$, in the domain of fixed target experiments.  The ratio
of the cross sections for transversely and longitudinally
polarized vector mesons will give a measure of $h_{1}$ provided one
does not have contamination by the higher twist process where the
photon is transversely polarized.  The interesting fact here is
that one does not need to polarize the proton.  Now $h_{1}$ involves
a matrix element off-diagonal in helicity (in the limit
$t=0=x_{1}-x_{2}$).  So in an inclusive experiment we have to polarize
the protons if we are to measure $h_{1}$.  But in our process, one
of the protons in the matrix element is in the final-state.  To
get the cross section we square the matrix element and sum over
all spin states for the outgoing proton.  Thus the off-diagonal
nature of the matrix element is compatible with an unpolarized
cross section (as regards the proton).

\subsection{Production of pseudo-scalar mesons}

Exclusive pion production involves the helicity parton densities.
So it should not be suppressed at large $x$ compared to vector
meson production.  But it should be much smaller at small $x$.

A number of predictions can be made for ratios of cross sections
of different mesons, if some approximations are made\footnote{
    Note that the predictions made in the preprint version of
    this paper were based on incorrect reasoning.
}.  These are
that the meson wave functions are SU(3) symmetric, that the
strange quark helicity density $\Delta s$ is small, and that the
helicity distribution of the up and down quarks are approximately
equal and opposite: $\Delta d\approx -\Delta u$ (this follows from the
observation
that $F_{2}$ for the deuteron is small and the assumption that this
same property is valid for the off-diagonal parton densities).

Using the SU(3) wave functions and these approximations for the
parton densities, it can be verified that
\begin{eqnarray}
    \frac {d\sigma (e+p \to  \eta +p)/dt}{d\sigma (e+p \to  \pi ^{0}+p)/dt}
    &\approx &
    \frac {1}{3} \left(
         \frac {2\Delta u_{V} - \Delta d_{V}}{2\Delta u_{V} + \Delta d_{V}}
      \right)^{2}
    \approx  3,
\nonumber\\
    \frac {d\sigma (e+p \to  \eta +p)/dt}{d\sigma (e+n \to  \eta +n)/dt}
    &\approx &
      \left(
         \frac {2\Delta u_{V} - \Delta d_{V}}{2\Delta d_{V} - \Delta u_{V}}
      \right)^{2}
    \approx  1,
\nonumber\\
    \frac {d\sigma (e+p \to  \pi ^{0}+p)/dt}{d\sigma (e+n \to  \pi ^{0}+n)/dt}
    &\approx &
      \left(
         \frac {2\Delta u_{V} + \Delta d_{V}}{2\Delta d_{V} + \Delta u_{V}}
      \right)^{2}
    \approx  1.
\end{eqnarray}
Here $\Delta u_{V}=\Delta u-\Delta \bar u$ and $\Delta d_{V}=\Delta d-\Delta
\bar d$.


\section{Conclusions}
\label{sec:conclusions}

We have proved a factorization theorem for exclusive
meson production in high $Q$ electroproduction.  The level of the
proof is comparable to that for the classic inclusive hard
scattering processes, like Drell-Yan.  An important consequence
is that higher-order corrections can be systematically calculated
in powers of $\alpha _{s}(Q)$.

We have derived new results that the theorem applies to large $x$
as well as to small $x$, and that it applies to the production of
all mesons, and not just vector mesons.  Thus we are able to
treat the process
\begin{equation}
    \gamma ^{*} + p \to  \pi ^{+} + n,
\end{equation}
for example.

In addition, we have shown that the theorem applies separately to
the case of production of transversely polarized vector mesons.
In that case we probe the $h_{1}$ or transversity distribution.
Although we expect this case to be suppressed at small $x$, we
see no reason for a suppression at large $x$.  This process then
provides an interesting new method to measure $h_{1}$, admittedly
the off-diagonal version.  An important consideration is that it
is not necessary to have any polarization information about the
proton, unlike the situation when one measures $h_{1}$ in inclusive
scattering.

The proof applies only to the case of that the virtual photon
that induces the scattering has longitudinal polarization. The
treatment of the same process with transversely polarized photons
appears to be a much harder problem in QCD, with a definite power
suppression.

An important question that needs further study is to understand
how much predictive power there is in the theorem.  As always
with perturbative QCD, the problem is that physical quantities
are represented in terms of parton densities etc which we are
unable to calculated perturbatively.  If we had predictions for
the non-perturbative quantities, we would have complete
predictions for the cross-sections.  But at the present state of
the art, we only have models for the non-perturbative quantities,
and very little that can be regarded as QCD predictions from
first principles.  Only for the perturbative quantities, the hard
scattering and the evolution kernels, do we possess a systematic
method of calculation within QCD.

Now, for ordinary inclusive processes, we able to measure the parton
densities from a limited set of processes at one energy and then
predict many other processes at all energies that allow the hard
scattering to be perturbative.  The reason that this is
straightforward is that the parton densities are functions of just
one longitudinal variable, and that the deep-inelastic structure
functions depend on a corresponding variable, $x$.  Indeed, with
the lowest order hard scattering, the structure functions are
just simple linear combinations of parton densities.  Obvious
generalizations of these remarks apply to other processes, in
hadron-hadron scattering, for example.  An immediate consequence
is that it is possible to make many real predictions from QCD for
inclusive hard scattering.  (Of course, practical limitations
arise from uncalculated higher order corrections and from
substantial experimental errors.)

But the situation is totally different for our process of elastic
meson production.  The cross section is function of one
momentum-fraction variable, but we have total of three such
variables in the factorization formula.  It is not so obvious
that we can measure the non-perturbative quantities, even in
principle.

At small $x$, the situation is better, since the parton densities
are dominated by exchange of vacuum quantum numbers: we have
a normal Regge limit. To the extent that there are universal
Regge trajectories, we get a relation between the power laws for
the $x$ dependence in our process and in ordinary deep-inelastic
scattering.  Since there is an integral over a longitudinal
momentum fraction there need not be an exact relation between the
off-diagonal parton densities and the diagonal ones probed in
ordinary deep-inelastic scattering.  Thus we may not be able to
get precise quantitative information on the diagonal gluon
density, particularly as regards the normalization.  Our hope is
that by some kind of Regge factorization we could say that the
two parton densities differ by some kind of Regge vertex, and
that since this Regge vertex would be probed at large virtuality,
we might be able to calculate it.

In the leading $\ln x$ approximation, the leading non-diagonal
terms are in fact computable in terms of the diagonal parton
densities (in the limit $t \to  0$). Similarly, after evolution in
$Q^{2}$, the non-diagonal terms come dominantly from the calculated
evolution kernel, rather than from the non-diagonal terms in the
initial distribution.

Our proof of the theorem also applies to charge exchange
scattering.  Then the generalized parton densities are
off-diagonal in flavor.  They are related by an isospin
transformation to non-singlet parton densities (at non-zero
momentum transfer).  There should therefore be some possibilities
to improve the phenomenology of the ordinary non-singlet quark
densities from an analysis of processes like
\begin{equation}
   \gamma ^{*} + p \to  \rho ^{+} + n .
\end{equation}

Our analysis also has direct implications for scattering off nuclei
implying that color transparency phenomena should be present for
exclusive production of leading mesons.  We leave the discussion
of this subject to a separate paper.

{\em Note added in proof:}
In fact, the ``off-diagonal parton distributions'' that we use
were actually introduced long ago in Ref.\ \cite{BALO},
where the diffractive production of the Z boson in DIS
was considered. Subsequently there is a long history \cite{OFPD},
including the previously cited paper by Balitsky and Braun
\cite{Balitsky.Braun}.

\section*{Acknowledgments}

This work was supported in part by the U.S. Department of Energy
under Grant Nos.\ DE-FG02-90ER-40577 and DE-FG02-93ER40771, and
by the Binational Science Foundation under Grant No.\ 9200126.
JCC would like to thank CERN for support during part of the
writing of this paper.


\end{document}